\documentclass[twocolumn]{bmcart}

\usepackage{amsthm,amsmath}
\RequirePackage[numbers]{natbib}
\RequirePackage{hyperref}
\usepackage[utf8]{inputenc} %unicode support
\usepackage{amssymb} % math, it is always a good idea
\usepackage{xspace} % for i.e., etc. abbreviations
\usepackage{cleveref}
\usepackage{acronym}
\usepackage{multirow,booktabs,arydshln} % provide dashline
\usepackage[inline]{enumitem} % provide inline enumerate
\usepackage{pifont}% http://ctan.org/pkg/pifont
\usepackage[percent]{overpic} % provide \put on figures
\usepackage{subfigure}
\usepackage{ulem} % provide \sout for strikeout

\usepackage{graphicx}

\startlocaldefs

\newcommand{\cmark}{\ding{51}} % checkmark
\newcommand{\xmark}{\ding{55}} % x mark
\newcommand{\ang}[1]{\ensuremath{#1^\circ}} % angle in degree

\newcommand{\codeLibrary}[1]{\texttt{#1}}

\newcommand{\dEchorate}{\texttt{dEchorate}}
\newcommand{\voicehome}{\texttt{VoiceHome}}
\newcommand{\mirage}{\texttt{MIRaGE}}
\newcommand{\butReverb}{\texttt{BUT-Reverb}}

\newcommand{\linkCode}{\href{www.github.com/Chutlhu/dEchorate}{github}}%\textcolor{gray}{\texttt{www.github.com/Chutlhu/dEchorate}}}
\newcommand{\linkData}{\href{www.zenodo.com/SOMETHING}{zenodo}}%\textcolor{gray}{\texttt{www.zenodo.com/SOMETHING}}}

\newcommand{\linkPyrirtools}{\href{https://github.com/maj4e/pyrirtool}{\codeLibrary{pyrirtools}}}
\newcommand{\linkPeakutils}{\href{https://bitbucket.org/lucashnegri/peakutils/}{\codeLibrary{peakutils}}}
\newcommand{\linkPyroomacoustics}{\href{https://github.com/LCAV/pyroomacoustics}{\codeLibrary{pyroomacoustics}}}
\newcommand{\linkSpeechmetrics}{\href{https://github.com/aliutkus/speechmetrics/}{\codeLibrary{speechmetrics}}}

\newcommand{\RTsixty}{\ensuremath{\mathtt{RT}_{60}}}
\newcommand{\bIPS}{\ensuremath{\mathtt{bIPS}}}
\newcommand{\MDS}{\ensuremath{\mathtt{MDS}}}
\newcommand{\dMDS}{\ensuremath{\mathtt{dMDS}}}
\newcommand{\dcMDS}{\ensuremath{\mathtt{dcMDS}}}
\newcommand{\DP}{\ensuremath{\mathtt{DP}}}
\newcommand{\DS}{\ensuremath{\mathtt{DS}}}

\newcommand{\MVDRretfLate}{\ensuremath{\mathtt{MVDR\text{-}ReTF\text{-}Late}}}
\newcommand{\MVDRrakeLate}{\ensuremath{\mathtt{MVDR\text{-}Rake\text{-}Late}}}
\newcommand{\SNR}{\ensuremath{\mathtt{SNR}}}

\newcommand{\bb}[1]{\ensuremath{\mathbb{#1}}}

\newcommand{\mics}{\ensuremath{\mathbf{x}}}
\newcommand{\flts}{\ensuremath{\mathbf{h}}}
\newcommand{\nses}{\ensuremath{\mathbf{n}}}
\newcommand{\weights}{\ensuremath{\mathbf{w}}}

\newcommand{\src}{\ensuremath{s}}

\makeatletter
\DeclareRobustCommand\onedot{\futurelet\@let@token\@onedot}
\def\@onedot{\ifx\@let@token.\else.\null\fi\xspace}
\makeatother
\newcommand{\eg}{\textit{e.g\onedot}}
\newcommand{\ie}{\textit{i.e\onedot}}

\newcommand{\wrt}{w.r.t\onedot}

\makeatletter
\AtBeginDocument{%
  \renewcommand*\AC@hyperlink{%
    \ifAC@starred
      \expandafter\@secondoftwo
    \else
      \expandafter\hyperlink
    \fi
  }%
}
\makeatother

\acrodefplural{TOA}[TOAs]{Times of Arrival}
\acrodefplural{TDOA}[TDOAs]{Time Differences of Arrival}
\acrodefplural{DOA}[DOAs]{Directions of Arrival}

\makeatletter
\AtBeginDocument{%
  \renewcommand*{\AC@hyperlink}[2]{%
    \begingroup
      \hypersetup{hidelinks}%
      \hyperlink{#1}{#2}%
    \endgroup
  }%
}
\makeatother

\endlocaldefs

\begin{document}

%%% Start of article front matter
\begin{frontmatter}

\begin{fmbox}
\dochead{Research}

% https://asmp-eurasipjournals.springeropen.com/spatial-audio
\title{dEchorate: a Calibrated Room Impulse Response Database for  Echo-aware Signal Processing}

%%%%%%%%%%%%%%%%%%%%%%%%%%%%%%%%%%%%%%%%%%%%%%
%%                                          %%
%% Enter the authors here                   %%
%%                                          %%
%% Specify information, if available,       %%
%% in the form:                             %%
%%   <key>={<id1>,<id2>}                    %%
%%   <key>=                                 %%
%% Comment or delete the keys which are     %%
%% not used. Repeat \author command as much %%
%% as required.                             %%
%%                                          %%
%%%%%%%%%%%%%%%%%%%%%%%%%%%%%%%%%%%%%%%%%%%%%%

\author[
  addressref={aff1},                   % id's of addresses, e.g. {aff1,aff2}
  % corref={aff1},                       % id of corresponding address, if any
  noteref={n1},                        % id's of article notes, if any
  email={diego.di-carlo@inria.fr}   % email address
]{\inits{D.D.C.}\fnm{Diego} \snm{Di Carlo}}
\author[
  addressref={aff2},
]{\inits{A.D.}\fnm{Pinchas} \snm{Tandeitnik}}
\author[
  addressref={aff3},
]{\inits{A.D.}\fnm{C\'edric} \snm{Foy}}
\author[
  addressref={aff4},
]{\inits{A.D.}\fnm{Antoine} \snm{Deleforge}}
\author[
  addressref={aff1},
]{\inits{N.B.}\fnm{Nancy} \snm{Bertin}}
\author[
  addressref={aff2},
]{\inits{A.D.}\fnm{Sharon} \snm{Gannot}}

%%%%%%%%%%%%%%%%%%%%%%%%%%%%%%%%%%%%%%%%%%%%%%
%%                                          %%
%% Enter the authors' addresses here        %%
%%                                          %%
%% Repeat \address commands as much as      %%
%% required.                                %%
%%                                          %%
%%%%%%%%%%%%%%%%%%%%%%%%%%%%%%%%%%%%%%%%%%%%%%

\address[id=aff1]{%                           % unique id
  \orgname{Univ Rennes, Inria, CNRS, IRISA},
  \cny{France}
}
\address[id=aff2]{%
 \orgname{Faculty of Engineering, Bar-Ilan University, Ramat-Gan, 5290002},
 \cny{Israel}
}
\address[id=aff3]{%
  \orgname{UMRAE, Cerema, Univ. Gustave Eiffel, Ifsttar, Strasbourg, 67035},
  \cny{France}
}
\address[id=aff4]{%
  \orgname{Universit\'e de Lorraine, Inria,  CNRS, LORIA, F-54000 Nancy},
  \cny{France}
}

%%%%%%%%%%%%%%%%%%%%%%%%%%%%%%%%%%%%%%%%%%%%%%
%%                                          %%
%% Enter short notes here                   %%
%%                                          %%
%% Short notes will be after addresses      %%
%% on first page.                           %%
%%                                          %%
%%%%%%%%%%%%%%%%%%%%%%%%%%%%%%%%%%%%%%%%%%%%%%

\begin{artnotes}
%\note{Sample of title note}     % note to the article
\note[id=n1]{The first author performed the work
while at Bar-Ilan University.} % note, connected to author
\end{artnotes}

% \end{fmbox}% comment this for two column layout

%%%%%%%%%%%%%%%%%%%%%%%%%%%%%%%%%%%%%%%%%%%%%%%
%%                                           %%
%% The Abstract begins here                  %%
%%                                           %%
%% Please refer to the Instructions for      %%
%% authors on https://www.biomedcentral.com/ %%
%% and include the section headings          %%
%% accordingly for your article type.        %%
%%                                           %%
%%%%%%%%%%%%%%%%%%%%%%%%%%%%%%%%%%%%%%%%%%%%%%%

\begin{abstractbox}

\begin{abstract} % abstract
% \parttitle{First part title} %if any
This paper presents dEchorate: a new database of measured multichannel Room Impulse Responses (RIRs) including annotations of early echo timings and 3D positions of microphones, real sources and image sources under different wall configurations in a cuboid room.
These data provide a tool for benchmarking recent methods in \textit{echo-aware} speech enhancement, room geometry estimation, RIR estimation, acoustic echo retrieval, microphone calibration, echo labeling and reflectors estimation.
The database is accompanied with software utilities to easily access, manipulate and visualize the data as well as baseline methods for echo-related tasks.
\end{abstract}

%%%%%%%%%%%%%%%%%%%%%%%%%%%%%%%%%%%%%%%%%%%%%%
%%                                          %%
%% The keywords begin here                  %%
%%                                          %%
%% Put each keyword in separate \kwd{}.     %%
%%                                          %%
%%%%%%%%%%%%%%%%%%%%%%%%%%%%%%%%%%%%%%%%%%%%%%


% MSC classifications codes, if any
%\begin{keyword}[class=AMS]
%\kwd[Primary ]{}
%\kwd{}
%\kwd[; secondary ]{}
%\end{keyword}

\end{abstractbox}
\end{fmbox}% uncomment this for two column layout

\end{frontmatter}

%%%%%%%%%%%%%%%%%%%%%%%%%%%%%%%%%%%%%%%%%%%%%%%%
%%                                            %%
%% The Main Body begins here                  %%
%%                                            %%
%% Please refer to the instructions for       %%
%% authors on:                                %%
%% https://www.biomedcentral.com/getpublished %%
%% and include the section headings           %%
%% accordingly for your article type.         %%
%%                                            %%
%% See the Results and Discussion section     %%
%% for details on how to create sub-sections  %%
%%                                            %%
%% use \cite{...} to cite references          %%
%%  \cite{koon} and                           %%
%%  \cite{oreg,khar,zvai,xjon,schn,pond}      %%
%%                                            %%
%%%%%%%%%%%%%%%%%%%%%%%%%%%%%%%%%%%%%%%%%%%%%%%%

%%%%%%%%%%%%%%%%%%%%%%%%% start of article main body
% <put your article body there>
\section{Introduction}\label{sec:intro}

\begin{figure}
    \centering
    % \includegraphics[width=0.95\linewidth]{figures/rir_model_empty.png}
    % \caption{Depiction of the different components of a room impulse response as they relate to sound propagation.}
    \includegraphics[width=0.95\linewidth]{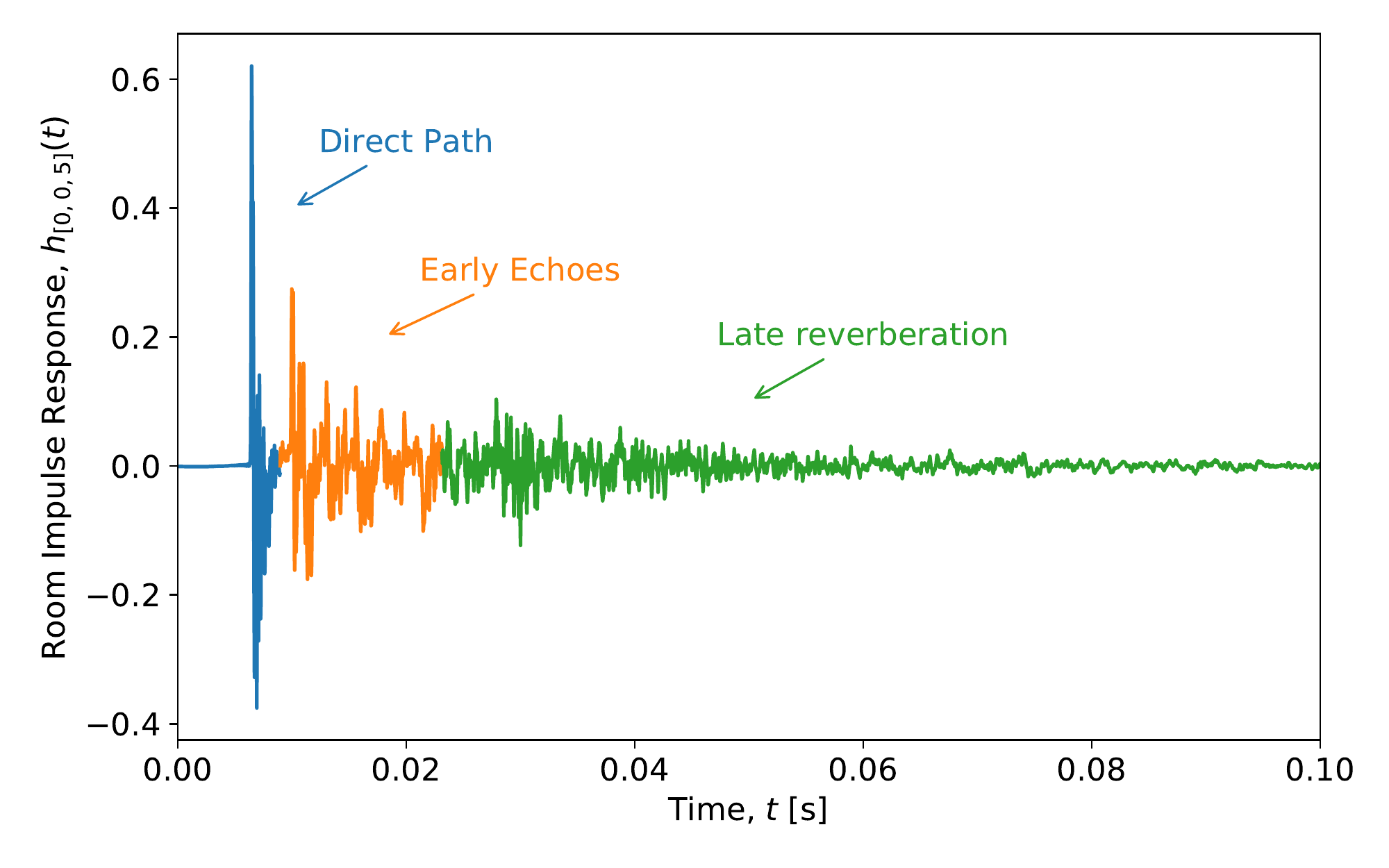}
    \caption{Depiction of a measured room impulse response from the database.}
    \label{fig:rir}
\end{figure}

% Sound Propagation
When sound travels from a source to a microphone in a indoor space, it interacts with the environment by being delayed and attenuated due to the distance; and reflected, absorbed and diffracted due to the surfaces.
The \ac{RIR} represents this phenomenon as a linear and causal time-domain filter.
As depicted in~\Cref{fig:rir}, \acp{RIR} are commonly subdivided into 3 parts:
the \textit{direct-path}, corresponding to the line-of-sight propagation; the \textit{early echoes}, stemming from few disjoint reflections on the closest reflectors; and the \textit{late reverberation} comprising the dense accumulation of later reflections and \textit{scattering} effects.

The late reverberation is indicative of the environment size and reverberation time, producing the so-called \textit{listener envelopment}, \ie, the degree of immersion in the sound field~\cite{griesinger1997psychoacoustics}.
In contrast, the direct path and the early echoes carry precise information on the 
%source content and the (NOT on the source content?)
scene's geometry, such as the position of the source and room surfaces relative to the receiver position~\cite{kuttruff2009room}, and on the surfaces' reflectivity.
Such relation is well explained by the \ac{ISM} \cite{allen1979image}, in which the echoes are associated with the contribution of virtual sound sources lying outside the real room.
Therefore, one may consider early echoes as ``spatialized'' copies of the source signal, whose \acp{TOA} are related to the source and reflector positions.

% Echo-aware processing
Based on this idea, so-called \textit{echo-aware} methods have been introduced a few decades ago, where \textit{matched filters} (or \textit{rake receivers}) are used to constructively sum the sound reflections \cite{flanagan1993spatially,jan1995matched,affes1997signal} and build beamformers achieving much better sound qualities \cite{gannot2001signal}.
These methods have recently regained interested as manifested by the European project SCENIC~\cite{annibale2011scenic} and the UK research project S$^3$A\footnote{\url{http://www.s3a-spatialaudio.org/}}.
Later, a few studies showed that knowing the properties of a few early echoes could boost the performance of typical indoor audio inverse problems such as \ac{SE} \cite{dokmanic2015raking,kowalczyk2019raking}, sound source localization \cite{ribeiro2010turning,salvati2016sound,dicarlo2019mirage,daniel2020time} and separation \cite{asaei2014structured,leglaive2016multichannel,scheibler2017separake,remaggi2019modeling}, and speaker verification~\cite{al2019early}.

Another fervent area of research spanning transversely the audio signal processing field is estimating the room geometry blindly from acoustic signals~\cite{antonacci2012inference,dokmanic2013acoustic,crocco2017uncalibrated,remaggi2017acoustic}.
As recently reviewed by Crocco \etal{} in \cite{crocco2017uncalibrated}, end-to-end \ac{RooGE} involves a number of subtasks:
\ac{RIR} estimation, peak picking, microphones calibration, echo labeling and reflectors' position estimation. As interesting applications, these methods have been recently used in active setting (\ie, knowing the transmitted signals) on unmanned aerial vehicles (UAVs, a.k.a. drones) \cite{jensen2019method,boutin2020drone} and on mobile-phones \cite{shih2019phone}.
The lowest common denominator of all these tasks is \ac{AER}, that is, estimating the properties of early echoes, such as their \acp{TOA} and energies. The former problem is typically referred to as \ac{TOA} estimation, or \ac{TDOA} estimation when the direct-path is taken as reference.

% % However most of them are on synthetic data or specialized data
% Despite the promising results achieved in all above-mentioned studies, most of ... \todoDDC{to be completed.}

% \usepackage{multirow}
\begin{table*}[t]
\caption{\label{tab:rir_db}Comparison between some existing RIR databases that account for early acoustic reflections. Receiver positions are indicated in terms of number of microphones per array times number of different positions of the array ($\sim$ stands for partially available information).
The read is invited to refer to \cite{szoke2018building,genovese2019blind} for more complete list of existing RIR datasets.
\protect\\$^{\dagger}$The dataset in \cite{remaggi2017acoustic} is originally intended for RooGE and further extended for (binaural) SE in \cite{remaggi2019modeling} with a similar setup.
\protect\\$^{\ddagger}$These datasets have been recorded in the same room.}
\small
\centering
\resizebox{1 \linewidth}{!}{%
\begin{tabular}{l|c|c|c|ccc|c|l|l}
\toprule

\multirow{2}{*}{Database Name} &
  \multicolumn{3}{c|}{Annotated} &
  \multicolumn{4}{c|}{Number of} &
  Key characteristics &
  Purpose \\
  &
  Pos. &
  Echoes &
  Rooms &
  \multicolumn{1}{c|}{RIRs} &
  \multicolumn{1}{c|}{Rooms} &
  \multicolumn{1}{c|}{Mic$\times$Pos.} &
  Src &
   &
   \\
\hline
\begin{tabular}[c]{@{}l@{}} Dokmani\'c \textit{et al.} \cite{dokmanic2013acoustic}\end{tabular} &
  \cmark  &
  $\sim$ &
  \multicolumn{1}{c|}{$\sim$} &
  \multicolumn{1}{c|}{15} &
  \multicolumn{1}{c|}{3} &
  \multicolumn{1}{c|}{5} &
  1 &
  Non shoebox rooms &
  RooGE \\ \hline
\begin{tabular}[c]{@{}l@{}} Crocco \textit{et al.} \cite{crocco2017uncalibrated}\end{tabular} &
  \cmark &
  $\sim$ &
  \multicolumn{1}{c|}{\cmark} &
  \multicolumn{1}{c|}{204} &
  \multicolumn{1}{c|}{1} &
  \multicolumn{1}{c|}{17} &
  12 &
  \begin{tabular}[c]{@{}l@{}}Accurate 3D calibration\\ Many mic and src positions\end{tabular} &
  RooGE \\
 \hline

 Remaggi \textit{et al.}~\cite{remaggi2017acoustic}$^{\dagger}$ &
  \cmark & % # noete pos?
  $\sim$  & % # note echo?
  \cmark & % # note room?
  \multicolumn{1}{c|}{$\sim$1.5k} & % # rirs
  \multicolumn{1}{c|}{4} & % # room
  \multicolumn{1}{c|}{48$\times$2} & % # mics
  4-24 & % # srcs
  \begin{tabular}[c]{@{}l@{}}
  Circural dense array
  \\Circular placement of sources
  \end{tabular} &
  \begin{tabular}[c]{@{}l@{}}RooGE\\ SE\end{tabular}
  \\
\hdashline
Remaggi \textit{et al.} \cite{remaggi2019modeling}$^{\dagger}$ &
  \cmark & % # noete pos?
  $\sim$  & % # note echo?
  \cmark & % # note room?
  \multicolumn{1}{c|}{$\sim$1.6k} & % # rirs
  \multicolumn{1}{c|}{4} & % # room
  \multicolumn{1}{c|}{
    \begin{tabular}[c]{@{}l@{}}
    48$\times$2
    \\+2$\times$2
    \end{tabular}
} & % # mics
  3-24 & % # srcs
  \begin{tabular}[c]{@{}l@{}}
    Circural dense array
    \\Binaural Recordings
    \end{tabular}
  &
  \begin{tabular}[c]{@{}l@{}}RooGE\\ SE\end{tabular}
  \\
\hline
%%%%%%%%%%%%%%%%%%%%%%%%%%%%%% BIU's
\texttt{BIU's Database}~\cite{hadad2014multichannel}$^{\ddagger}$ &
    \cmark&
    \xmark&
    \xmark&
    \multicolumn{1}{c|}{$\sim$1.8k} &
    \multicolumn{1}{c|}{3} &
    \multicolumn{1}{c|}{8$\times$3} &
    26  &
    \begin{tabular}[c]{@{}l@{}}
        Linear array with different spacing\\
        Circular placement of sources
    \end{tabular} &
    SE \\
\hline
%%%%%%%%%%%%%%%%%%%%%%%%%%%%%% BUTREVERB
\butReverb~\cite{szoke2018building} &
  \cmark &
  \xmark&
  $\sim$ &
  \multicolumn{1}{c|}{$\sim$1.3k} &
  \multicolumn{1}{c|}{8} &
  \multicolumn{1}{c|}{(2-10)$\times$6} &
  3-11 &
  \begin{tabular}[c]{@{}l@{}}
    Accurate metadata\\
    different device/arrays\\
    various rooms
  \end{tabular} &
  SE/ASR \\ \hline
%%%%%%%%%%%%%%%%%%%%%%%%%%%%%% VOICEHOME
\voicehome~\cite{bertin2019voice} &
   \cmark&
   \xmark&
   \xmark&
  \multicolumn{1}{c|}{188} &
  \multicolumn{1}{c|}{12} &
  \multicolumn{1}{c|}{8$\times$2} &
  7-9  &
   Various rooms, real homes &
  SE/ASR \\ \hline
%%%%%%%%%%%%%%%%%%%%%%%%%%%%%% MIRAGE
\mirage~\cite{cmejla2021mirage}$^{\ddagger}$ &
    \cmark&
    \xmark&
    \xmark&
    \multicolumn{1}{c|}{371k} &
    \multicolumn{1}{c|}{3} &
    \multicolumn{1}{c|}{5$\times$6} &
    25 (+ 4104)  &
    \begin{tabular}[c]{@{}l@{}}
    4104 src. pos. in a dense grid\\
    different acoustic rooms\end{tabular} &
    SE/ASR \\ \hline
%%%%%%%%%%%%%%%%%%%%%%%%%%%%%% DECHORATE
\dEchorate$^{\ddagger}$ &
  \textbf{\cmark} &
  \textbf{\cmark} &
  \multicolumn{1}{c|}{\textbf{\cmark}} &
  \multicolumn{1}{c|}{$\sim$1.8k} &
  \multicolumn{1}{c|}{11} &
  \multicolumn{1}{c|}{5$\times$6} &
  6 &
  \begin{tabular}[c]{@{}l@{}}
    Accurate echo annotation\\
    different surface absorptions
   \end{tabular} &
  \begin{tabular}[c]{@{}l@{}}RooGE\\ SE/ASR\end{tabular}
  \\

\bottomrule
\end{tabular}
}
\end{table*}

As listed in \cite{szoke2018building} and in \cite{genovese2019blind}, a number of recorded \acp{RIR} corpora are freely available online, each of them meeting the demands of certain applications. \Cref{tab:rir_db} summarizes the main characteristics of some of them.
One can broadly identify two main classes of echo-aware \ac{RIR} datasets in the literature: \ac{SE}/\ac{ASR}-oriented datasets, \eg~\cite{szoke2018building,bertin2019voice,cmejla2021mirage}, and \ac{RooGE}-oriented datasets, \eg{}~\cite{dokmanic2013acoustic,crocco2017uncalibrated,remaggi2017acoustic}.
The former regards acoustic echoes as highly correlated interfering sources coming from close reflectors, such as a table in a meeting room or a near wall. This typically presents a challenge in estimating the correct source's \ac{DOA} with further consequences in \ac{DOA}-based enhancement algorithm, \eg, beamformers. 
Although this factor is taken into account, such datasets lack proper annotation of these echoes in the \acp{RIR} or the absolute position of objects inside the room.
The latter group typically features design choices, such as microphones scattered across the room, which are not suitable for \ac{SE} applications. Indeed, these typically involve compact or ad hoc arrays.
The main common drawback of these datasets in that they cannot be easily used for other tasks than the ones which they are designed for.

To bypass the complexity of recording and annotating real \ac{RIR} datasets, acoustic simulators based on the \ac{ISM} are extensively used instead~\cite{gaultier2017vast,kim2017generation,perotin2018crnn,kim2017generation,dicarlo2020blaster}.
While such data are more versatile, simpler and quicker to obtain, they fail to fully capture the complexity and richness of real acoustic environments.
Due to this, methods trained, calibrated, or validated on them may fail to generalize to real conditions, as will be shown in this paper.
Interestingly, in the context of learning-based blind room volume estimation, the authors of \cite{genovese2019blind} combined multiple real and synthetic \ac{RIR} datasets in order to find a balance between number of training data and realism.

A good echo-oriented \ac{RIR} dataset should include a variety of environments (room geometries and surface materials), of microphone placings (close to or away from reflectors, scattered or forming ad-hoc arrays) and, most importantly, precise annotations of the scene's geometry and echo timings in the \acp{RIR}.
Moreover, in order to be versatile and used in both \ac{SE} and \ac{RooGE} applications, geometry and timing annotations should be fully consistent.
Such data are difficult to collect since it involves precise measurements of the positions and orientations of all the acoustic emitters, receivers and reflective surfaces inside the environment with dedicated planimetric equipment.

To fill this gap, we present the \dEchorate{} dataset: a fully calibrated multichannel \ac{RIR} database with accurate annotation of the geometry and echo timings in different configurations of a cuboid room with varying wall acoustic profiles.
The database currently features 1800 annotated \acp{RIR} obtained from 6 arrays of 5 microphones each, 6 sound sources and 11 different acoustic conditions.
All the measurements were carried out at the acoustic lab at Bar-Ilan University following a consolidated protocol previously established for the realization of two other multichannel \acp{RIR} databases: the BIU's Impulse Response Database \cite{hadad2014multichannel} gathering \acp{RIR} of different reverberation levels sensed by uniform linear arrays (ULAs); and \mirage~\cite{cmejla2021mirage} providing a set of measurements for a source placed on a dense position grid.
The \dEchorate{} dataset is designed for AER with linear arrays, and is more generally aimed at analyzing and benchmarking \ac{RooGE} and echo-aware signal processing methods on real data.
In particular, it can be used to assess robustness against the number of reflectors, the reverberation time, additive spatially-diffuse noise and non-ideal frequency and directive characteristics of microphone-source pairs and surfaces in a controlled way.
Due to the amount of data and recording conditions, it could also be used to train machine learning models or as a reference to improve \ac{RIR} simulators.
The database is accompanied with a Python toolbox that can be used to process and visualize the data, perform analysis or annotate new datasets.

\begin{figure*}
    \centering
    \includegraphics[trim={0 0 0 10em},clip,width=0.95\textwidth]{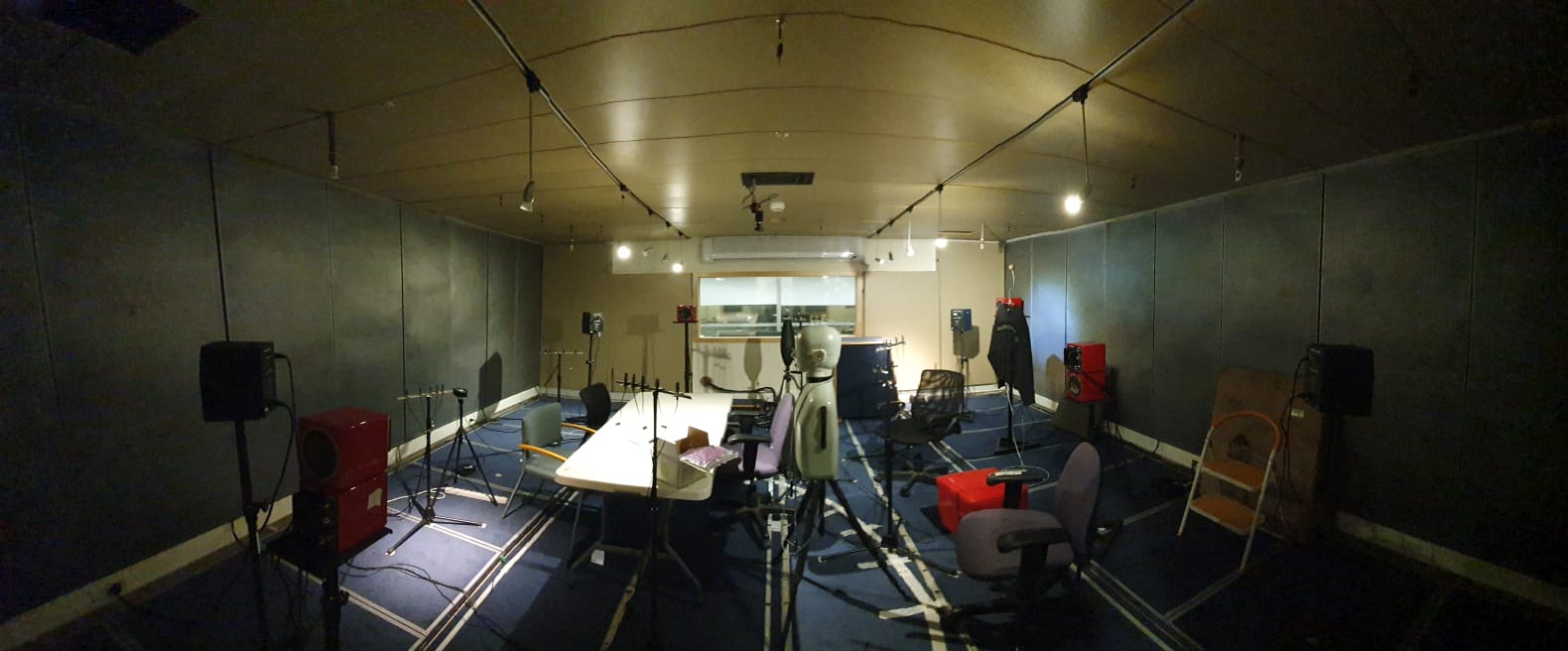}
    \caption{Broad-view picture of the acoustic lab at Bar-Ilan university.}
    \label{fig:fornitures}
\end{figure*}

The remainder of the paper is organized as follows.
\Cref{sec:description} describes the construction and the composition of the dataset, while \Cref{sec:analysis} provides an overview of the data, studying the variability of typical acoustic parameters.
To validate the data, in \Cref{sec:applications} two echo-aware application are presented, one in speech enhancement and one is room geometry estimation.
Finally, in~\Cref{sec:conclusion} the paper closes with the conclusions and and offers leads for work.
\section{Database Description}\label{sec:description}

\begin{table}[]
    \caption{\label{tab:room_equipment} Measurement and recording equipment.}

    \centering
    \small
    \begin{tabular}{ll}
        \toprule
         Loudspeakers   & (directional, direct) $4 \times$ Avanton\\
                        & (directional, indirect) $2 \times$ Avanton\\
                        & (omnidirectional) $1 \times$ B\&G\\
                        & (babble noise) $4 \times$ 6301bx Fostex\\
         \hline
         Microphones    & $30 \times$ AKG CK32\\
         Array          & $6 \times$ nULA (5 mics each, handcrafted)\\
         \hline
         A/D Converter  & ANDIAMO.MC\\
         \hline
         Indoor Positioning & Marvelmind Starter Set HW v4.9\\
         \bottomrule
    \end{tabular}
\end{table}

\subsection{Recording setup}
The recording setup is placed in a cuboid room with dimension 6 m $\times$ 6 m $\times$ 2.4 m.
The 6 facets of the room (walls, ceiling, floor) are covered by acoustic panels allowing controllable reverberation time (\RTsixty).
We placed 4 directional loudspeakers (direct sources) facing the center of the room and 30 microphones mounted on 6 static linear arrays parallel to the ground.
An additional channel is used for the loop-back signal, which serves to compute the time of emission and detect errors.
Each loudspeaker and each array is positioned close to one of the walls in such a way that the source of the strongest echo can be easily identified.
Moreover, their positioning was chosen to cover a wide distribution of source-to-receiver distances, hence, a wide range of \acp{DRR}.
Further, 2 more loudspeakers were positioned pointing towards the walls (indirect sources).
This was done to study the case of early reflections being stronger than the direct-path.

Each linear array consists of 5 microphones with non-uniform inter-microphone spacings of $[4, 5, 7.5, 10]$ cm\footnote{%
    \footnotesize
that is, $[-12.25, -8.25, -3.25, 3.25, 13.25]$ cm \wrt the barycenter}.
Hereinafter we will refer to these elements as \acp{nULA}.
% Each array is steered towards a different vertical edge of the room for calibration and reproducibility purposes.

\begin{figure}[t]
    \centering
    \includegraphics[width=0.95\linewidth]{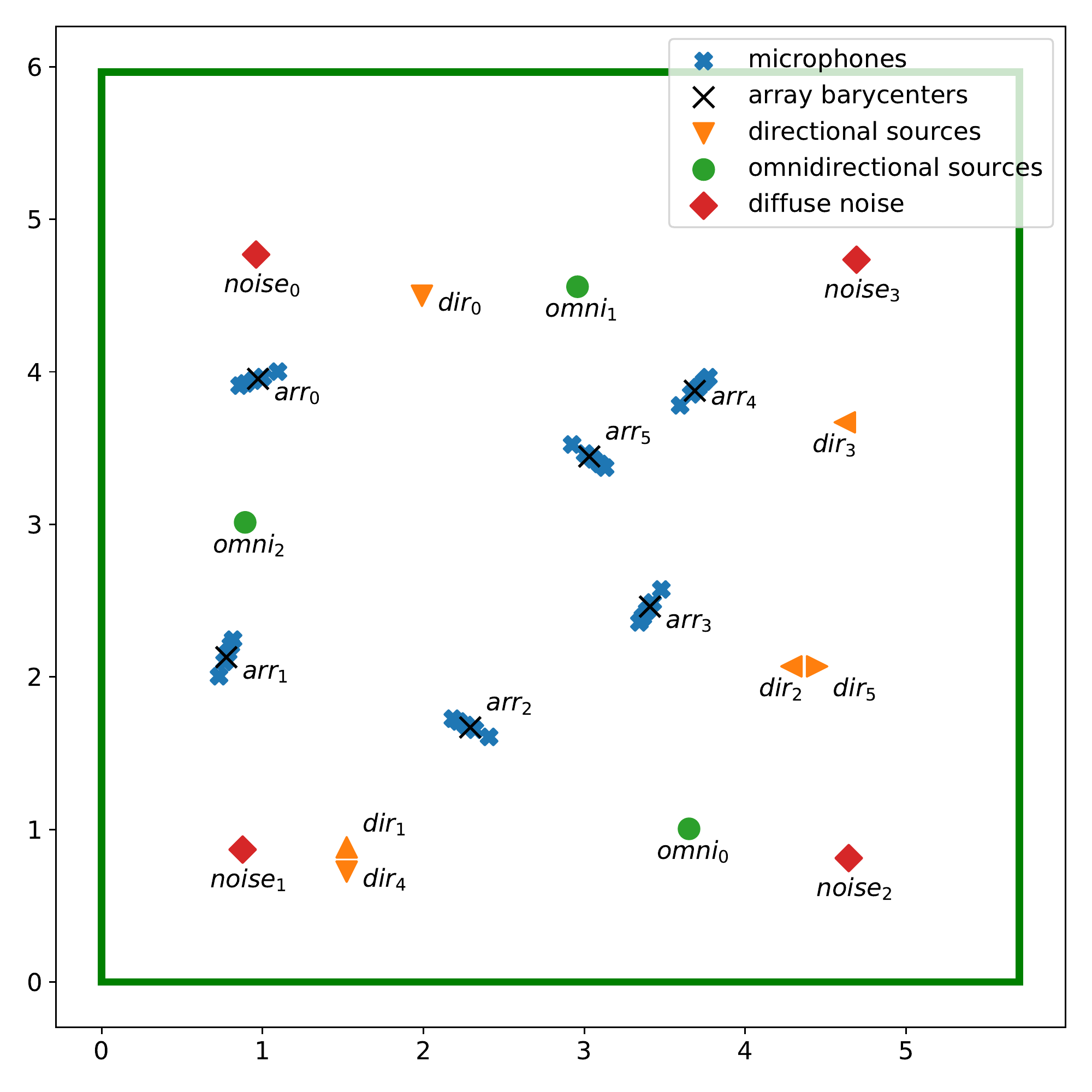}
    \caption{Illustration of the recording setup - top view.}
    \label{fig:2D}
\end{figure}

\begin{figure*}[h]
    \hfill
    \subfigure{
        \includegraphics[width=0.31\textwidth]{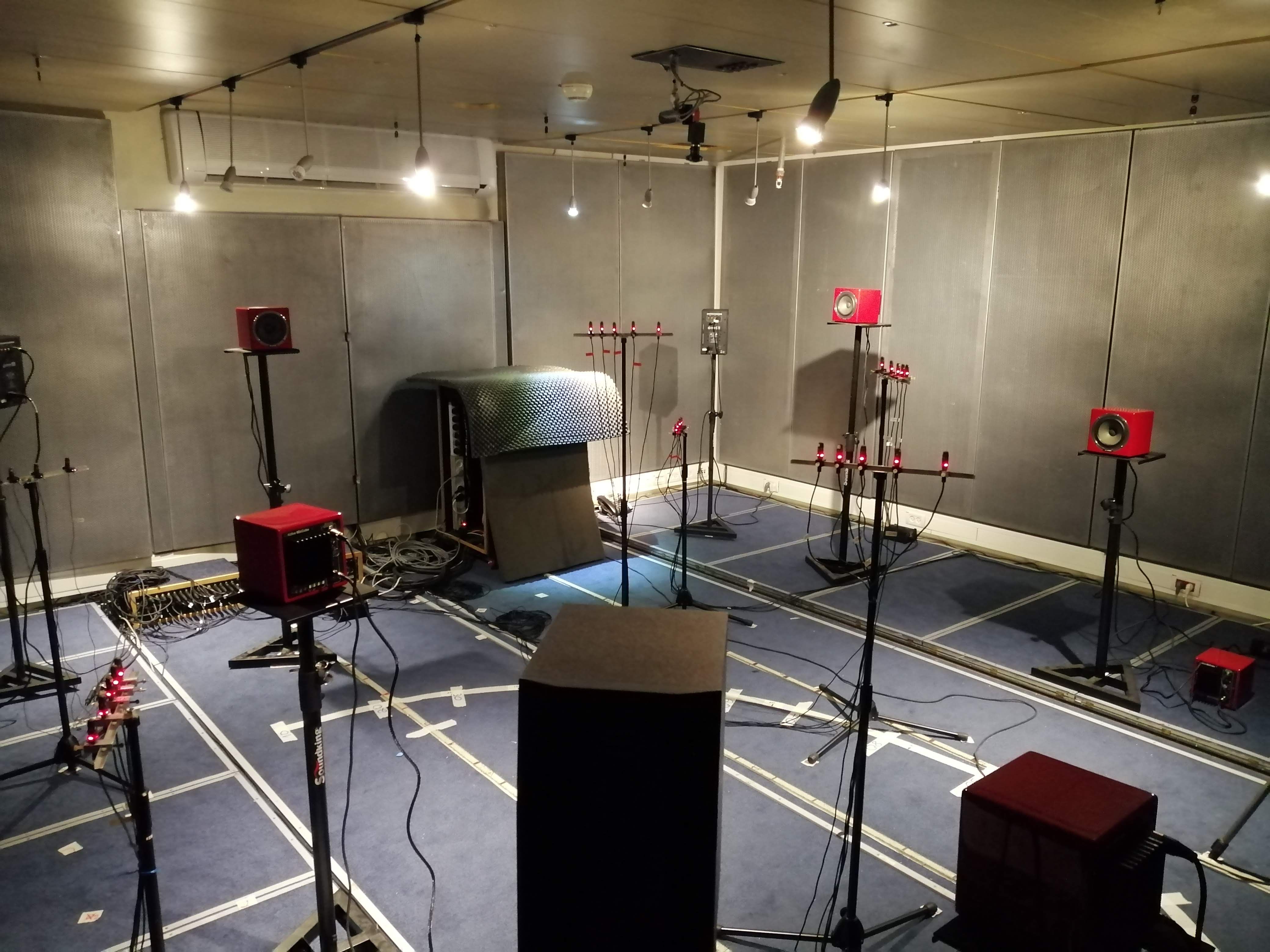}}
    \hfill
    \subfigure{
        \includegraphics[width=0.31\textwidth]{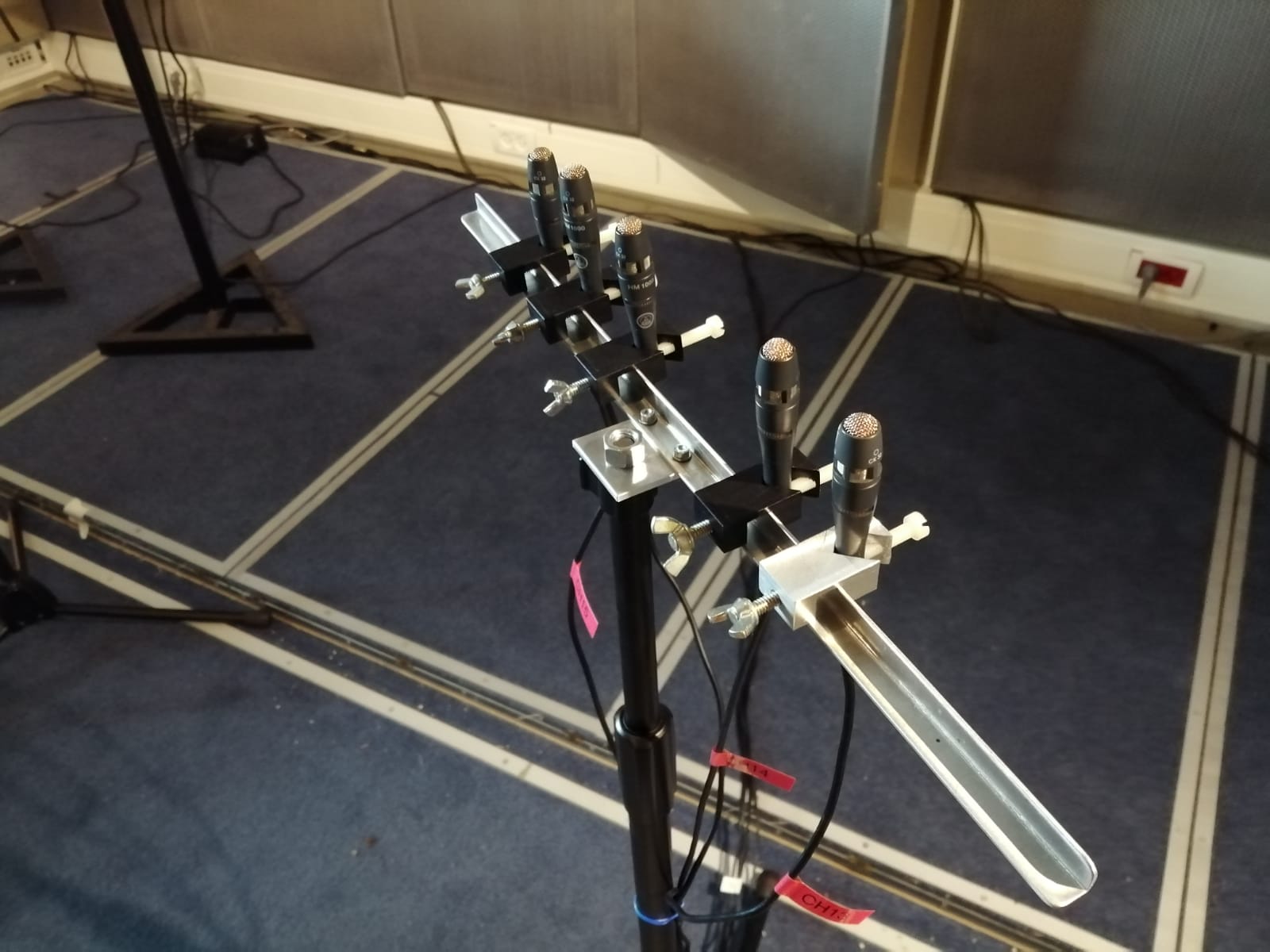}}
    \hfill
    \subfigure{
        \includegraphics[width=0.31\textwidth]{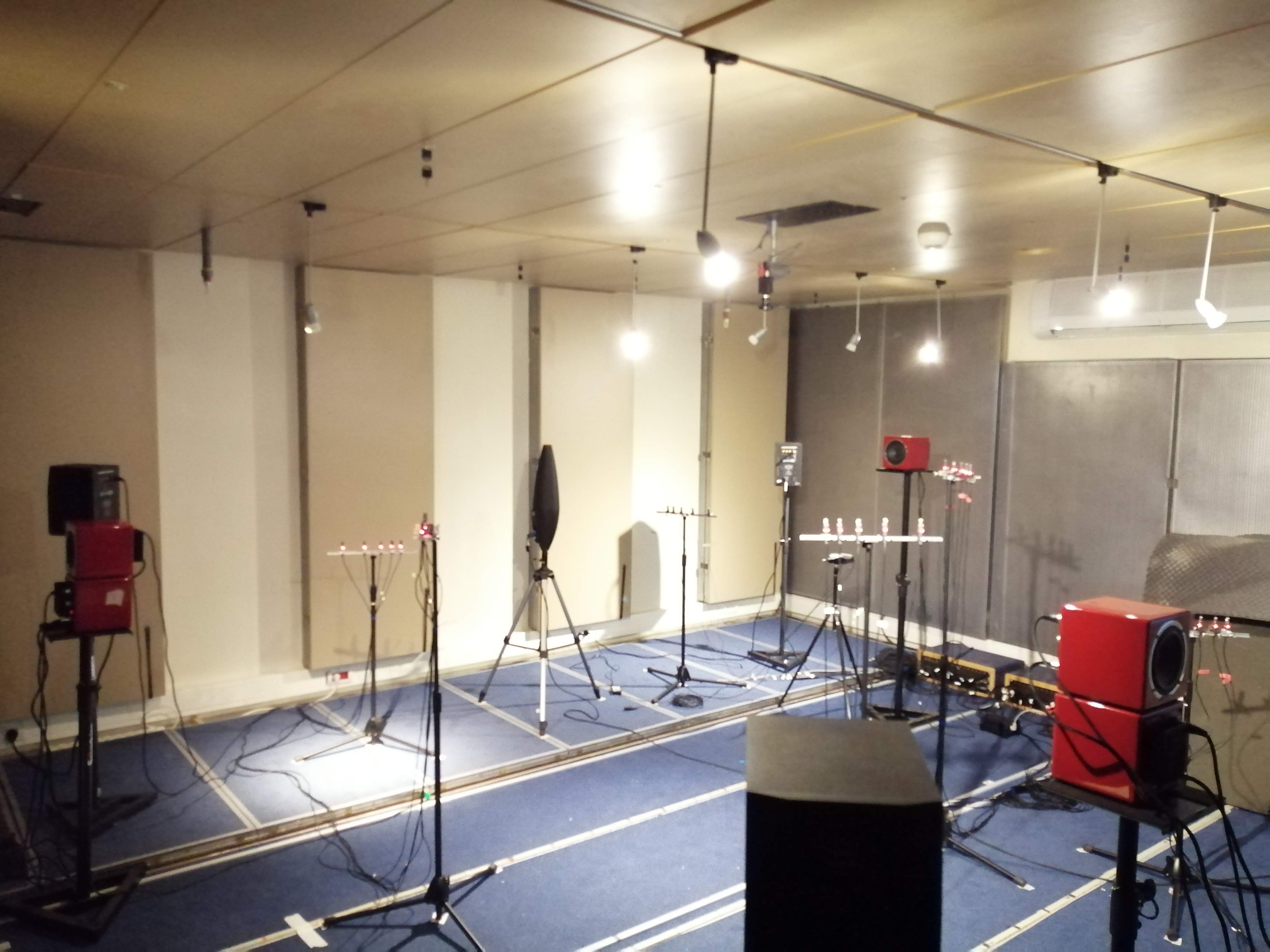}}
    \caption{Picture of the acoustic lab. From left to right: the overall setup, one microphone array, the setup with revolved panels.}
    \label{fig:setup}
\end{figure*}

\subsection{Measurements}
The main feature of this room is the possibility to change the acoustic profile of  each of its facets by flipping double-sided panels with one reflective (made of Formica Laminate sheets) and one absorbing face made of perforated panels filled with rock-wool). 
A complete list of the materials of the room is available in \Cref{app:materials}.
This allows to achieve diverse values of $\RTsixty$ that range from 0.1 to almost 1~second.
In this dataset, the panels of the floor were always kept absorbent.

Two types of measurement sessions were considered, namely, \textit{one-hot} and \textit{incremental}.
For the first type, a single facet was placed in reflective mode while all the others were kept absorbent.
For the second type, starting from fully-absorbent mode, facets were progressively switched to reflective one after the other until all but the floor are reflective, as shown in~\Cref{tab:wallcoding}.
The dataset features an extra recording session.
For this session, office furnitures (chairs, coat-hanger and a table) were positioned in the room to simulate a typical meeting room with chairs and tables (see~\Cref{fig:fornitures}).
Theses recordings may be used to assert the robustness of echo-aware methods in a more realistic scenario

\begin{table}[]
\small
\caption{\label{tab:wallcoding} Surface coding in the dataset: each binary digit indicates if the surface is absrobent ($\mathtt{0}$, \xmark ) or reflective ($\mathtt{1}$, \cmark).}

\begin{tabular}{p{.5cm}p{1.1cm}|p{.5cm}p{.5cm}p{.5cm}p{.5cm}p{.5cm}c}
\toprule
& Surfaces: & Floor & Ceil & West & South & East & North\\
\hline

\multicolumn{1}{c}{\multirow{5}{*}{\rotatebox{90}{one-hot}}} & $\mathtt{010000}$                   & \xmark & \cmark & \xmark & \xmark & \xmark & \xmark \\
% & \multicolumn{1}{c|}{$\vdots$}  & \multicolumn{6}{c}{$\ddots$} \\
& $\mathtt{001000}$ & \xmark & \xmark & \cmark & \xmark & \xmark & \xmark \\
& $\mathtt{000100}$ & \xmark & \xmark & \xmark & \cmark & \xmark & \xmark \\
& $\mathtt{000010}$ & \xmark & \xmark & \xmark & \xmark & \cmark & \xmark \\
& $\mathtt{000001}$ & \xmark & \xmark & \xmark & \xmark & \xmark & \cmark \\
\hdashline
\multicolumn{1}{c}{\multirow{6}{*}{\rotatebox{90}{incremental}}} & $\mathtt{000000}$                    & \xmark & \xmark & \xmark & \xmark & \xmark & \xmark \\
& $\mathtt{010000}$ & \xmark & \cmark & \xmark & \xmark & \xmark & \xmark \\
& $\mathtt{011000}$ & \xmark & \cmark & \cmark & \xmark & \xmark & \xmark \\
& $\mathtt{011100}$ & \xmark & \cmark & \cmark & \cmark & \xmark & \xmark \\
& $\mathtt{011110}$ & \xmark & \cmark & \cmark & \cmark & \cmark & \xmark \\
& $\mathtt{011111}$ & \xmark & \cmark & \cmark & \cmark & \cmark & \cmark \\
\hdashline
\multicolumn{1}{c}{\multirow{1}{*}{\rotatebox{90}{f.}}} 
& $\mathtt{010001}^{*}$ & \xmark & \cmark & \xmark & \xmark & \xmark & \cmark \\
\bottomrule
\end{tabular}
\end{table}

For each room configuration and loudspeaker, three different excitation signals were played and recorded in sequence: chirps, white noise and speech utterances.
The former consists in a repetition of 3 \ac{ESS} signals of duration 10 seconds and frequency range from 100 Hz to 14 kHz interspersed with 2 seconds of silence.
Such frequency range was chosen to match the characteristics of the loudspeakers.
To prevent rapid phase changes and ``popping'' effects, the signals were linearly faded in and out over 0.2 seconds with a Tuckey taper window.\footnote{\label{fn:pyrir}%
    \footnotesize The code to generate the reference signals and to process them is available together with the data.
    The code is based on the \linkPyrirtools{} Python library}
Second, 10 seconds bursts of white noise and 3 anechoic speech utterances from the \ac{WSJ} dataset~\cite{paul1992design} were played in the room.
Through all recordings, at least 40 dB of sound dynamic range compared to the room silence was asserted, and a room temperature of $\ang{24} \pm \ang{0.5}$C and $80\%$ relative humidity were registered. In these conditions the speed of sounds is $c_\text{air} = 346.98 $ m/s.
In addition, 1 minute of \textit{room tone} (\ie, silence) and 4~minutes of diffuse babble noise were recorded for each session. The latter was simulated by transmitting different chunks of the same single-channel babble noise recording from additional loudspeakers facing the four corners of the room.

All microphone signals were synchronously acquired and digitally converted to 48~kHz with 32~bit/sample using the equipment listed in~\Cref{tab:room_equipment}.
The polarity of each microphone was recorded by clapping a book in the middle of the room and their gain is corrected using the room tone. 

Finally, \acp{RIR} are estimated with the \ac{ESS} technique~\cite{farina2007advancements} where an exponential time-growing frequency sweep is using as probe signal. Then, the \ac{RIR} is estimated by devolving the microphone signal, implemented as division in the frequency domain (The authors used the same code mentioned in \Cref{fn:pyrir}).
% the signal at a microphone listening to an \ac{ESS} source is deconvolved by division in the frequency domain.

\subsection{Dataset annotation}\label{subsec:annotation}

\subsubsection{RIRs annotation}
The objective of this database is to feature annotations in the ``geometrical space'', namely the microphone, facet and source positions,  that are \textit{fully consistent} with annotations in the ``signal space'', namely the echo timings within the \acp{RIR}.
This is achieved as follows:
\begin{enumerate}[label=(\roman*)]
    \item \label{it:decharate:ips} First, the ground-truth positions of the array and source centres are acquired via a Beacon indoor positioning system ($\bIPS$).
    This system consists in 4 stationary bases positioned at the corners of the ceiling and a movable probe used for measurements which can be located within errors of $\pm2$~cm.

    \item \label{it:decharate:not} The estimated \acp{RIR} are superimposed on synthetic \acp{RIR} computed with the \acf{ISM} from the geometry obtained in the previous step.
    A Python GUI\footnote{\footnotesize This GUI is available in the dataset package.} (showed in~\Cref{fig:labelling_tools}), is used to manually tune a peak finder and label the echoes corresponding to found peaks, that is, annotate their timings and their corresponding image source position and room facet label.

    \item \label{it:decharate:mds} By solving a simple \acf{MDS} problem \cite{dokmanic2015relax,crocco2016estimation,plinge2016acoustic}, refined microphone and source positions are computed from echo timings.
    The non-convexity of the problem is alleviated by using a good initialization (obtained at the previous step), by the high SNR of the measurements and, later, by including additional image sources in the formulation.
    The prior information about the arrays' structures reduced the number of variables of the problem, leaving the 3D positions of the sources and of the arrays' barycenters in addition to the arrays' tilt on the azimuthal plane.

    \item \label{it:decharate:lat} By employing a multilateration algorithm \cite{beck2008exact}, where the positions of one microphone per array serve as anchors and the \acp{TOA} are converted into distances, it is possible to localize image sources alongside the real sources.
    This step will be further discussed in~\Cref{sec:applications}.
\end{enumerate}
Knowing the geometry of the room, in step \ref{it:decharate:ips} we were able to initially guess the position of the echoes in the \ac{RIR}. Then, by iterating through steps \ref{it:decharate:not}, \ref{it:decharate:mds} and \ref{it:decharate:lat}, the position of the echoes are refined to be consistent under the \ac{ISM}.

The final geometrical and signal annotation was chosen as a compromise between the $\bIPS$ measurements and the $\MDS$ output.
While the former ones are noisy but consistent with the scene's geometry, the latter ones match the \acp{TOA} but not necessarily the physical world.
In particular, geometrical ambiguities such as global rotation, translation and up-down flips were observed.
Instead of manually correcting this error, we modified the original problem from using only the direct path distances ($\dMDS$) to considering the  image sources' \ac{TOA} of the ceiling as well in the cost function ($\dcMDS$).
\Cref{tab:res_mds} shows numerically the \textit{mismatch} (in cm) between the geometric space (defined by the $\bIPS$ measurements) and the signal space (the one defined by the echo timings, converted to cm based on the speed of sound).
To better quantify it, we introduce here a \textit{\ac{GoM}} metric: it measures the fraction of (first-order) echo timings annotated in the \acp{RIR} matching the annotation produced by the geometry within a threshold.
Including the ceiling information, $\dcMDS$ produces a geometrical configuration which has a small mismatch (0.4~cm on average, 1.86~cm max) in both the signal \textit{and} geometric spaces with a $98.1\%$ matching all the first order echoes within a 0.5~ms threshold (\ie, the position of all the image sources within about 17~cm error).
It is worth noting that the $\bIPS$ measurements produce a significantly less consistent annotation with respect to the signal space.

\begin{table}[]
\centering
\caption{\label{tab:res_mds} Mismatch between geometric measurements and signal measurements in terms of maximum (Max.), average (Avg.) and standard deviation (Std) of absolute mismatch in centimeters.
The goodness of match (GoM) between the signal and geometrical measurements is reported as the fraction of matching echo timings for different thresholds in milliseconds.}

\begin{tabular}{lllll}
\toprule
& Metrics        & $\bIPS$               & $\dMDS$          & $\dcMDS$          \\
\midrule
\multicolumn{1}{c}{\multirow{2}{*}{\rotatebox{90}{\scriptsize Geom.}}}
&   Max.             & 0            & $6.1$         & $1.07$        \\
&   Avg.$\pm$Std.    & 0            & $1.8\pm1.4$    & $0.39\pm0.2$  \\
% \rule{0pt}{0.05em}\\
\midrule
\multicolumn{1}{c}{\multirow{2}{*}{\rotatebox{90}{\scriptsize Signal}}}
&   Max.          & $5.86$         & $1.20$         & $1.86$       \\
&   Avg.$\pm$Std. & $1.85\pm 1.5$  & $0.16\pm0.2$   & $0.41\pm0.3$ \\
% \rule{0pt}{0.05em}\\
\midrule
\multicolumn{1}{c}{\multirow{3}{*}{\rotatebox{90}{\scriptsize Mismatch}}}
&  GoM (0.5 ms)   & $97.9 \%$      & $93.4 \%$      & $98.1 \%$ \\
&  GoM (0.1 ms)   & $26.6 \%$      & $44.8 \%$      & $53.1 \%$ \\
&  GoM (0.05 ms)  & $12.5 \%$      & $14.4 \%$      & $30.2 \%$ \\
% \rule{0pt}{0.05em}\\
\bottomrule
\end{tabular}
\end{table}

\subsubsection{Other tools for RIRs annotation}
Finally, we would like to add that the following tools and techniques were found useful in annotating the echoes.

\paragraph{The ``skyline'' visualization} consists in presenting the intensity of multiple \acp{RIR} as an image, such that the wavefronts corresponding to echoes can be highlighted \cite{baba2018b}.
Let $h_{n}(l)$ be an \ac{RIR} from the database, where $l = 0, \ldots, L-1$ denotes sample index and $n = 0, \ldots, N-1$ is an arbitrary indexing of all the microphones for a fixed room configuration.
Then, the \textit{skyline} is the visualization of the $L \times N$ matrix $\mathbf{H}$ created by stacking column-wise $N$ normalized \textit{echograms}\footnote{
    \footnotesize
    The echogram is defined either as the absolute value or as the squared value of the \ac{RIR}.
}, that is 
\begin{equation}
    \mathbf{H}_{l, n} = \mid h_{n}(l) \mid / \max{ \mid{h_{n}(l)}\mid},
\end{equation}
where $\mid{\cdot}\mid$ denotes the absolute value.

\begin{figure*}
    \centering
    \includegraphics[trim={15em 15em 2em 0},clip,width=0.95\textwidth]{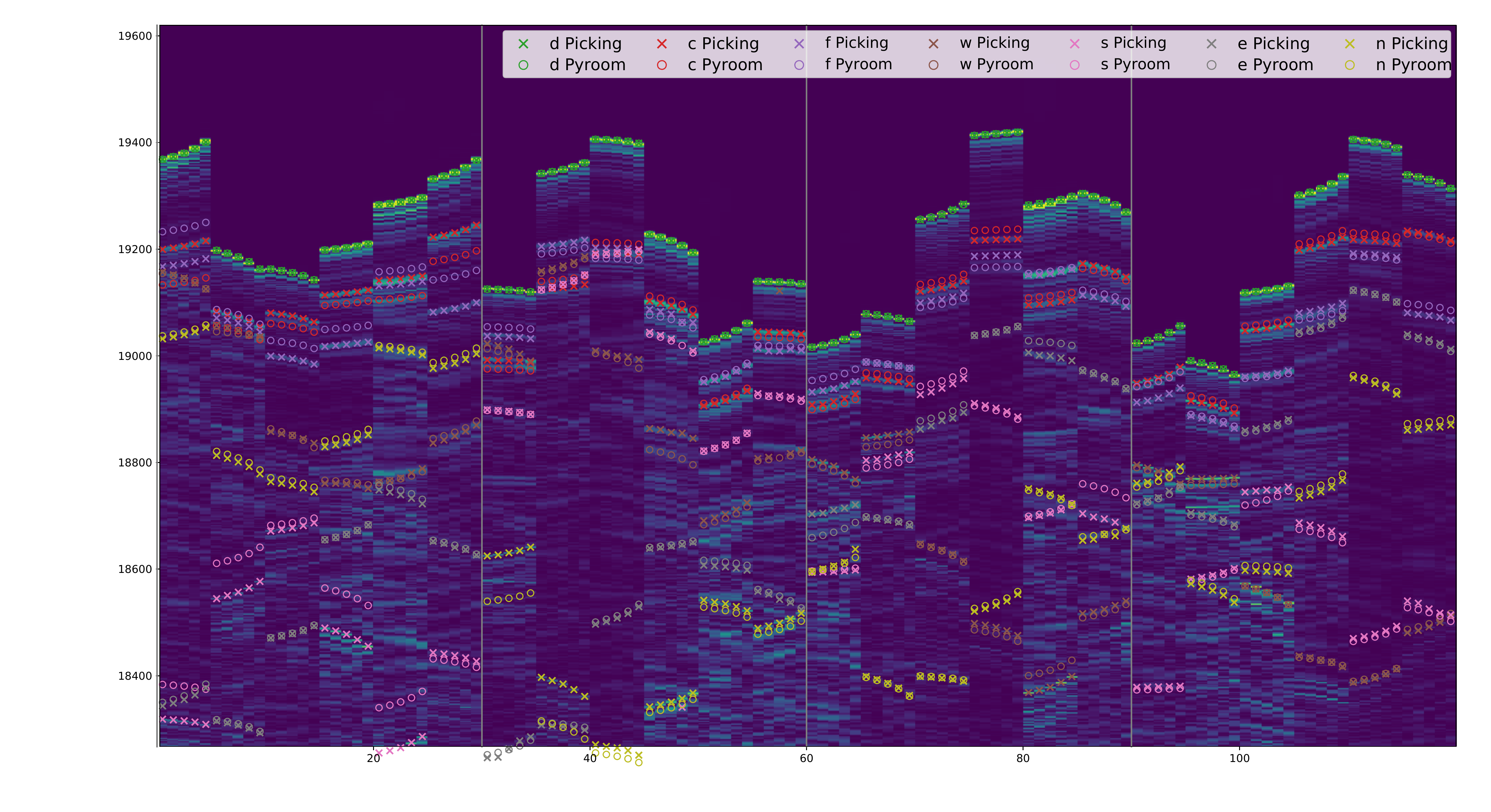}
    \caption{
        The \ac{RIR} skyline annotated with observed peaks ($\times$) together with their geometrically-expected position ($\circ{}$) computed with the Pyroomacoustic acoustic simulator.
        As specified in the legend, markers of different colors are used to indicate the room facets responsible for the reflection: direct path ($\mathtt{d}$), ceiling ($\mathtt{c}$), floor ($\mathtt{f}$), west wall ($\mathtt{w}$), $\dots$, north wall ($\mathtt{n}$).
    }\label{fig:skyline}
\end{figure*}

\Cref{fig:skyline} shows an example of skyline for 120 \acp{RIR} corresponding to 4 directional sources, 30 microphones and the most reflective room configuration, stacked horizontally, preserving the order of microphones within the arrays.
One can notice several clusters of 5 adjacent bins of similar color (intensity) corresponding to the arrivals at the 5 sensors of each \ac{nULA}.
Thanks to the usage of linear arrays, this visualization allowed us to identify both \acp{TOA} and their labeling.

\begin{figure*}[h]
    \centering
    \small
    \includegraphics[width=0.95\textwidth]{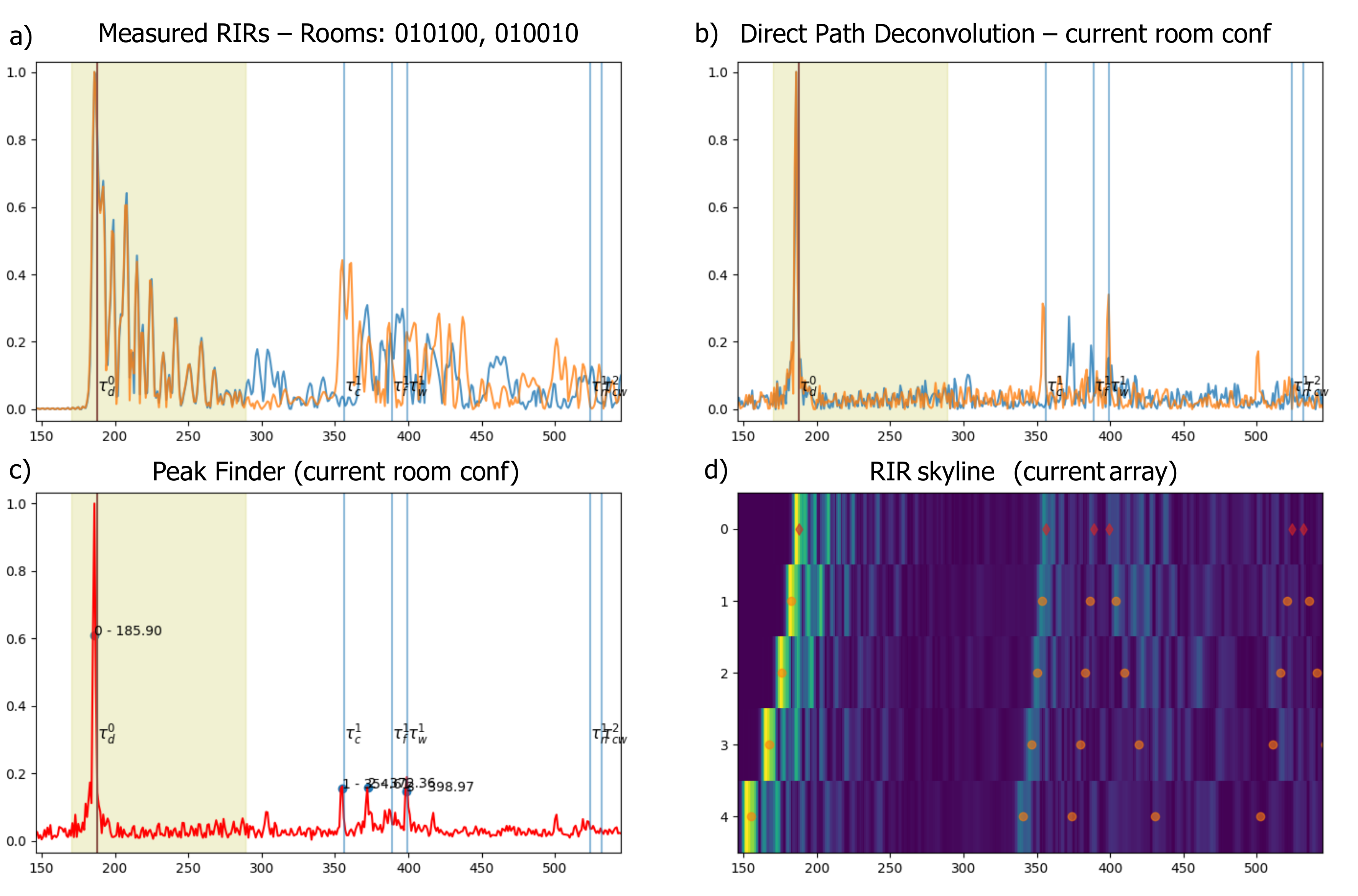}
    \caption{
        Detail of the GUI used to manually annotate the \acp{RIR}.
        For a given source and a microphone in an \ac{nULA},
        a) and b) each shows 2 \acp{RIR} for 2 different room configurations (blue and orange) before and after the direct path deconvolution.
        c) shows the results of the peak finder for one of the deconvolved RIRs, and d) is a detail on the \ac{RIR} skyline (See \Cref{fig:skyline}) on the corresponding \ac{nULA}, transposed to match the time axis.
    } \label{fig:labelling_tools}
\end{figure*}

\paragraph{Direct path deconvolution/equalization} was used to compensate for the frequency response of the source loudspeaker and microphone \cite{antonacci2012inference,eaton2016estimation}.
In particular, the direct path of the \ac{RIR} was manually isolated and used as an equalization filter to enhance early reflections from their superimposition before proceed with peak picking.
Each \ac{RIR} was equalized with its respective direct path.
As depicted in~\Cref{fig:labelling_tools}, in some cases this process was required for correctly identifying the underlying \acp{TOA}' peaks.

\paragraph{Different facet configurations} for the same geometry influenced the peaks' predominance in the \ac{RIR}, hence facilitating its echo annotation.
An example of \acp{RIR} corresponding to 2 different facet configurations is shown in~\Cref{fig:labelling_tools}: the reader can notice how the peak predominance changes for the different configurations.

\paragraph{An automatic peak finder} was used on equalized echograms $\bar{\eta}_{n}(l)$ to provide an initial guess on the peak positions.
In this work, peaks are found using the Python library \linkPeakutils{} whose parameters were manually tuned.

\subsection{Limitations of current annotation}
As stated in \cite{defrance2008finding}, we want to emphasize that annotating the correct \acp{TOA} of echoes and even the direct path in ``clean'' real \acp{RIR} is far from straightforward.
The peaks can be blurred out by the loudspeaker characteristics or the concurrency of multiple reflections.
Nevertheless, as showed in~\Cref{tab:res_mds}, the proposed annotation was found to be sufficiently consistent both in the geometric and in the echo/signal space.
Thus, no further refinement was done.
This database can be used as a first basis to develop better \ac{AER} methods which could be used to iteratively improve the annotation, for instance including  2$^\text{nd}$ order reflections.

\subsection{The \dEchorate~package}
The dataset comes with both data and code to parse and process it.
The data are presented in 2 modalities: the \texttt{raw} data, that is, the collection of recorded wave files, are organized in folders and can be retrieved by querying a simple database table; the \texttt{processed} data, which comprise the estimated \acp{RIR} and the geometrical and signal annotations, are organized in tensors directly importable in Matlab or Python (\textit{e.g.} all the \acp{RIR} are stored in a tensor of dimension $L \times I \times J \times D$, respectively corresponding to the RIR length in samples, the number of microphones, of sources and of room configurations).
\\Together with the data a Python package is available on the same website.
This includes wrappers, GUI, examples as well as the code to reproduce this study.
In particular, all the scripts used for estimating the \acp{RIR} and annotating them are available and can be used to further improve and enrich the annotation or as baselines for future works.
\section{Analysing the Data}\label{sec:analysis}
In this section we will illustrate some characteristics of the collected data in term of acoustic descriptors, namely the $\RTsixty$, the \ac{DRR} and the \ac{DER}. While the former two are classical acoustic descriptors used to evaluate \ac{SE} and \ac{ASR} technologies~\cite{eaton2015ace}, the latter is less common and used in strongly echoic situations~\cite{eargle1996characteristics,naylor2010speech}.

\subsection{Reverberation Time}
The $\RTsixty$ is the time required for the sound level in a room to decrease by 60 dB after the source is turned off, thus, it measures reverberation level. This value is one the most common acoustic descriptor for room acoustics. Besides, as reverberation affects detrimentally the performances of speech processing technologies, the robustness against $\RTsixty$ has become a common evaluation metric in \ac{SE} and \ac{ASR}.

\Cref{tab:rtsixty} reports estimated $\RTsixty(b)$ values per octave band $b\in \{ 500,1000,2000,4000 \}$ (Hz) for each of the room in the dataset. These values were estimated using the Schroeder's integration methods~\cite{schroeder1965new,chu1978comparison,xiang1995evaluation} in each octave band. For the octave bands centred at 125 Hz and 250 Hz, the measured \acp{RIR} did not exhibit sufficient power for a reliable estimation. This observation found confirmation in the frequency response provided by the loudspeakers' manufacturer, which decays exponentially from 300 Hz downwards.

Ideally, for the $\RTsixty$ to be reliably estimated, the Schroeder curve, \textit{i.e.} the log of the square-integrated, octave-band-passed \ac{RIR}, would need to feature a linear decay for 60 dB of dynamic range, which would occur in an ideal diffuse sound regime. However, such range is never observable in practice, due to the presence of noise and possible non-diffuse effects. Hence, a common technique is to compute, \textit{e.g.}, the $\mathtt{RT}_{10}$ on the range $[-5,-15]$ dB of the Schroeder curve and to extrapolate the $\RTsixty$ by multiplying it by 6. We visually inspected all the \acp{RIR} of the dataset corresponding to directional sources 1, 2 and 3, \textit{i.e.}, 90 \acp{RIR} in each of the 10 rooms. Then, two sets were created. Set $\mathcal{A}$ features all the Schroeder curves featuring linear log-energy decays allowing for reliable $\mathtt{RT}_{10}$ estimates.
Set $\mathcal{B}$ contains all the other curves. In practice, $49\%$ of the 3600 Schroeder curves were placed in the set $\mathcal{B}$. These mostly correspond to the challenging measurement conditions purposefully included in our dataset, \textit{i.e.}, strong early echoes, loudspeakers facing towards reflectors or receivers close to reflectors. Finally, the $\RTsixty$ value of each room and octave band was calculated from the median of $\mathtt{RT}_{10}$ corresponding to Schroeder curves in $\mathcal{A}$ only.

As can be seen in \Cref{tab:rtsixty}, obtained reverberation values are consistent with the room progressions described in \cref{sec:description}. Considering the 1000 Hz octave band, the $\RTsixty$ ranges from 0.14 s for the fully absorbent room ($\mathtt{000000}$) to 0.73 s for the most reflective room ($\mathtt{011111}$). When only one surfaces is reflective the $\RTsixty$ values remains around 0.19 s.

\begin{table*}[t!]
 \caption{\label{tab:rtsixty} Reverberation time per octave bands $\RTsixty(b)$ calculated in the 10 room configurations. For each coefficient, the number of corresponding Schroeder curves in $\mathcal{A}$ used to compute the median estimate is given in parentheses.}
 %\footnotesize
 \resizebox{1 \linewidth}{!}{%
 \begin{tabular}{l|cccccccccc}
    \toprule
    & Room 1 & Room 2 & Room 3 & Room 4 & Room 5 & Room 6 & Room 7 & Room 8 & Room 9 & Room 10 \\
    & $\mathtt{000000}$ & $\mathtt{011000}$  & $\mathtt{011100}$ & $\mathtt{011110}$  
    & $\mathtt{011111}$ & $\mathtt{001000}$  & $\mathtt{000100}$ & $\mathtt{000010}$  & $\mathtt{000001}$ & $\mathtt{010001}^*$ \\
    \midrule
    500 Hz  & 0.18 (11) & 0.40 (7) & 0.46 (20) & 0.60 (51) & 0.75 (48) & 0.22 (8) &	0.21 (5) &	0.21 (8) & 0.22 (7) & 0.37 (12) \\
    1000 Hz & 0.14 (62) & 0.33 (83) & 0.34 (86) & 0.56 (89) & 0.73 (90) & 0.19 (79) & 0.19 (74) & 0.18 (69) & 0.19 (70) & 0.26 (72) \\
    2000 Hz & 0.16 (65) & 0.25 (81) & 0.30 (86) & 0.48 (82) & 0.68 (88) & 0.18 (74) & 0.20 (64) &	0.18 (66) & 0.18 (67) & 0.24 (69) \\
    4000 Hz & 0.22 (15) & 0.25 (17) & 0.37 (22) & 0.55 (16) & 0.81 (29) & 0.22 (17) & 0.23 (12) & 0.26 (14) & 0.24 (18) & 0.28 (14) \\
    \bottomrule
 \end{tabular}}
\end{table*}

\subsection{Direct To Early and Reverberant Ratio}
In order to characterize an acoustic environment, it is common to provide the ratio between the energy of the direct and the indirect propagation paths. 
In particular, one can compute the so-called \ac{DRR} directly from a measured \ac{RIR} $h(l)$~\cite{eaton2015ace} as
\begin{equation}
    \mathtt{DRR} = 10 \log_{10} \frac{\sum_{l \in \mathcal{D}} h^2(l) }{\sum_{l \in \mathcal{R}} h^2l)} \quad [\text{dB}],
\end{equation}
where $\mathcal{D}$ denotes the time support comprising the direct propagation path (set to $\pm$120 samples around its time of arrival, blue part in \Cref{fig:rir}), and $\mathcal{R}$ comprises the remainder of the \ac{RIR}, including both echoes and late reverberation (orange and green parts in \Cref{fig:rir}).

Similarly, the \ac{DER} defines the ratio between the energy of the direct path and the early echoes only, that is,
\begin{equation}
    \mathtt{DER} = 10 \log_{10} \frac{\sum_{l \in \mathcal{D}} h^2(l) }{\sum_{l \in \mathcal{E}} h^2(l)} \quad [\text{dB}],
\end{equation}
where $\mathcal{E}$ is the time support of the early echoes only (green part in \Cref{fig:rir}).

Differently from the $\RTsixty$ which mainly describes the diffuse regime, both \ac{DER} and \ac{DRR} are highly dependent on the position of the source and receiver in the room. Therefore, for each room, wide ranges of these parameters were registered. For the loudspeakers facing the microphones, the \ac{DER} ranges from 2 dB to 6 dB in one-hot room configurations and from -2 dB to 6 dB in the most reverberant rooms.
Similarly, the \ac{DRR} has a similar trend featuring lower values, such as -2 dB in one-hot rooms and down to -7.5 dB for the most reverberant ones. A complete annotation of these metrics is available in the database.
\section{Using the Data}\label{sec:applications}
% In this section we exemplify the utilization of the database considering three possible use-cases: acoustic echo estimation, echo-aware spatial beamforming and room geometry estimation.
The dEchorate database
%with measured multichannel RIRs including annotations of early echoes is
%–finally–
is now used to investigate the performance of state-of-the-art methods on two echo-aware acoustic signal processing applications on both synthetic and measured data, namely, spatial filtering and room geometry estimation.

\subsection{Application: Echo-aware Beamforming}

Let $I$ microphones acquire to a single static point sound source, contaminated by noise sources.
In the short-time Fourier transform (STFT) domain, we stack the $I$ complex-valued microphone observations at frequency $f$ and time $t$ into a vector $\mics(f,t) \in \bb{C}^I$.
Let us denote $\src(f,t) \in \bb{C}$ and $\nses(f,t) \in \bb{C}^{I}$ the source signal and the noise signals at microphones, which are assumed to be statistically independent.
By denoting $\flts \in \bb{C}^I$ the Fourier transforms of the \acp{RIR}, the observed microphone signals in the STFT domain can be expressed a follows:
\begin{equation}
    \mics (f,t)  = \flts (f) \src(f,t) + \nses(f,t).
\end{equation}
Here, the STFT windows are assumed long enough so that the discrete convolution-to-multiplication approximation holds well.

Beamforming is one of the most widely used techniques for enhancing multichannel microphone recordings. The literature on this topic spans several decades of array processing and a recent review can be found in~\cite{gannot2017consolidated}.
In the frequency domain, the goal of beamforming is to estimate a set of coefficients $\weights(f) \in \bb{C}^{I}$ that are applied to $\mics(f,t)$, such that $s(f,t) \approx \weights^{H} \mics(f,t)$.
Hereinafter, we will consider only the \textit{distortionless} beamformers aiming at retrieving the clean target speech signal, as it is generated at the source position.
%As opposed to Multichannel Wiener Filtering approaches, beamformers are designed to keep the target signal distortionless. 

As mentioned throughout the paper, the knowledge of early echoes is expected to boost spatial filtering performances. However, estimating these elements is difficult in practice.
To quantify this, we compare \textit{echo-agnostic} and \textit{echo-aware} beamformers.
In order to study their empirical potential, we will evaluate their performance using both synthetic and measured data, as available in the presented dataset.

Echo-agnostic beamformers do not need any echo-estimation step: they either ignore their contributions, as in the direct-path delay-and-sum beamformer ($\DS$)~\cite{vantrees2004optimum}, or they consider coupling filters between pairs of microphones, called \acp{ReTF}~\cite{gannot2001signal}.
%Note that as opposed to \ac{RIR} estimation and \ac{AER}, \ac{ReTF} estimation is a non-blind problem.
Note that contrary to \acp{RIR}, there exist efficient methods to estimate \acp{ReTF} from multichannel recordings of unknown sources (see~\cite[Section VI.B]{gannot2017consolidated} for a review).
The \acp{ReTF} can then be naturally incorporated in powerful beamforming algorithms achieving speech dereverberation and noise reduction in static~\cite{schwartz2014multi} and dynamic scenarios~\cite{kodrasi2017evd}.
In this work, \acp{ReTF} are estimated using \ac{GEVD} method~\cite{markovich2009multichannel}, using the approach illustanted in~\cite{markovich2018performance}.

Echo-aware beamformers fall in the category of \textit{rake receivers}, borrowing the idea from telecommunication where an antenna \textit{rakes} (\ie, combines) coherent signals arriving from different propagation paths~\cite{flanagan1993spatially,jan1995matched,affes1997signal}.
To this end, they typically consider that for each \ac{RIR} $i$, the delays and frequency-independent attenuation coefficients of $R$ early echoes are known, denoted here as $\tau_i^{(r)}$ and $\alpha_i^{(r)}$. 
In the frequency domain, this translates into the following:
\begin{equation}\label{sec:appl:echomodel}
  \flts(f) = \left[ \sum_{r=0}^{R-1} \alpha_i^{(r)} \, \exp \left( 2\pi j f \tau_{i}^{(r)} \right) \right]_i,
\end{equation}
where $r = 0, \ldots, R - 1$ denotes the reflection order, 

Recently, these methods have been used for noise and interferer suppression in~\cite{dokmanic2015raking,scheibler2015raking} and for noise and reverberation reduction in~\cite{javed2016spherical,kowalczyk2019raking}.
The main limitation of these works is that echo properties, or alternatively the position of image sources, must be known \textit{a priori}.
Hereafter, we will assume these properties known by using the annotations of the dEchorate dataset, as described in~\Cref{subsec:annotation}. In particular, we will assume that the \acp{RIR} follow the echo model~(\ref{sec:appl:echomodel}) with $R = 4$, corresponding to the 4 strongest echoes.
Knowing the echo delays, the associated attenuation coefficients are retrieved from the \acp{RIR} using a simple maximum-likelihood approach, as in~\cite[Eq. 10]{condat2015cadzow}.

We evaluate the performance of both types of beamformers on the task of noise and late reverberation suppression.
Different \ac{MVDR} beamformers are considered,
assuming either spatially white noise (\ie, classical $\DS$ design), diffuse noise (\ie, the Capon filter) or diffuse noise \textit{plus} the late reverberation~\cite{schwartz2016joint}.
In the latter case, the late reverberation statistics are modeled by a spatial coherence matrix~\cite{kuster2012objective} weighted by the late reverberation power, which is estimated using the procedure described in~\cite{schwartz2016joint}.

Overall, the different \ac{RIR} models considered are direct propagation ($\DP$, \ie, ignoring echoes), multipath propagation ($\mathtt{Rake}$, \ie, using 4 known early echoes)~\cite{dokmanic2015raking,kowalczyk2019raking} or the full reverberant propagation ($\mathtt{ReTF}$)~\cite{gannot2001signal,markovich2018performance}.\Cref{tab:bf_design} summarizes the considered beamformers designs.

\begin{table}[h]
    \centering
    \caption{Summary of the considered beamformers. ``n.'' and ``lr.'' are used as short-hand for noise and late reverberation. (*) denotes echo-aware beamformers.}
    \label{tab:bf_design}
    \footnotesize
    \begin{tabular*}{\linewidth}{@{\extracolsep{\fill}}lll@{}}
    \toprule
    Acronym & Steering Vectors & Noise Model \\
    \midrule
    DS~\cite{vantrees2004optimum}             & Direct Path AOA     & Spatially white n.\\
    MVDR-DP~\cite{vantrees2004optimum}        & Direct Path AOA     & Diffuse n.\\
    MVDR-ReTF~\cite{gannot2001signal}         & \ac{ReTF}        & Diffuse n.\\
    MVDR-Rake*~\cite{dokmanic2015raking}      & 4 Echoes/chan.     & Diffuse n.\\
    MVDR-DP-Late~\cite{kowalczyk2019raking}   & Direct Path AOA     & Spat.ly white n.+lr.\\
    MVDR-ReTF-Late~\cite{markovich2018performance} & \ac{ReTF}         & Diffuse n. + lr.\\
    MVDR-Rake-Late*~\cite{kowalczyk2019raking} & 4 Echoes/chan.      & Diffuse n. + lr.\\
    \bottomrule
    \end{tabular*}
\end{table}

Performances of the different designs are compared on the task of enhancing a target speech signal in a 5-channel mixture using the \acp{nULA} in the \dEchorate{} dataset.
They are tested in scenarios featuring high reverberation and diffuse babble noise, appropriately scaled to pre-defined signal-to-noise ratios $\SNR \in \{0, 10, 20\}$.
Using the \dEchorate{} data, we consider the room configuration $\mathtt{011111}$ ($\RTsixty \approx 600 $ ms) and all possible combinations of (target, array) positions.
Both real and corresponding synthetic \acp{RIR} are used, which are then convolved with anechoic utterances from the \ac{WSJ} corpus~\cite{paul1992design} and corrupted by recorded diffuse babble noise. The synthetic \acp{RIR} are computed with the Python library \linkPyroomacoustics~\cite{scheibler2017pyroomacoustics}, based purely on the \ac{ISM}. Hence, on synthetic \acp{RIR}, the known echo timings perfectly match the components in their early part (no model mismatch).

The evaluation is conducted similarly to the one in~\cite{kowalczyk2019raking} where the following metrics are considered:
\begin{itemize}
    \item the \ac{iSNRR} in dB, computed as the difference between the input $\mathtt{SNRR}$ at the reference microphone and the $\mathtt{SNRR}$ at the filter output;
    \item the \ac{iSRMR} \cite{falk2010non} to measure dereverberation;
    \item the \ac{iPESQ} score~\cite{rix2001perceptual}
    to assess the perceptual quality of the signal and indirectly the amount of artifacts.
\end{itemize}
Implementations of the SRMR and \ac{PESQ} metrics are available in the Python library \linkSpeechmetrics.
Both the \ac{SNRR} and the \ac{PESQ} are relative metrics, meaning they require a target reference signal.
Here we consider the clean target signal as the dry source signal convolved with the early part of the \ac{RIR} (up to $R$-th echo) of the reference (first) microphone.
On the one hand, this choice numerically penalizes both direct-path-based and \ac{ReTF}-based beamformers, which respectively aim at extracting the direct-path signal and the full reverberant signal in the reference microphone.
On the other hand, considering only the direct path or the full reverberant signal would be equally unfair for the other beamformers.
Moreover, including early echoes in the target signal is perceptually motivated since they are known to contribute to speech intelligibility~\cite{bradley2003importance}.

\begin{figure*}[ht]
    \centering
    \includegraphics[trim={0 10 10 0},clip,width=0.95\textwidth]{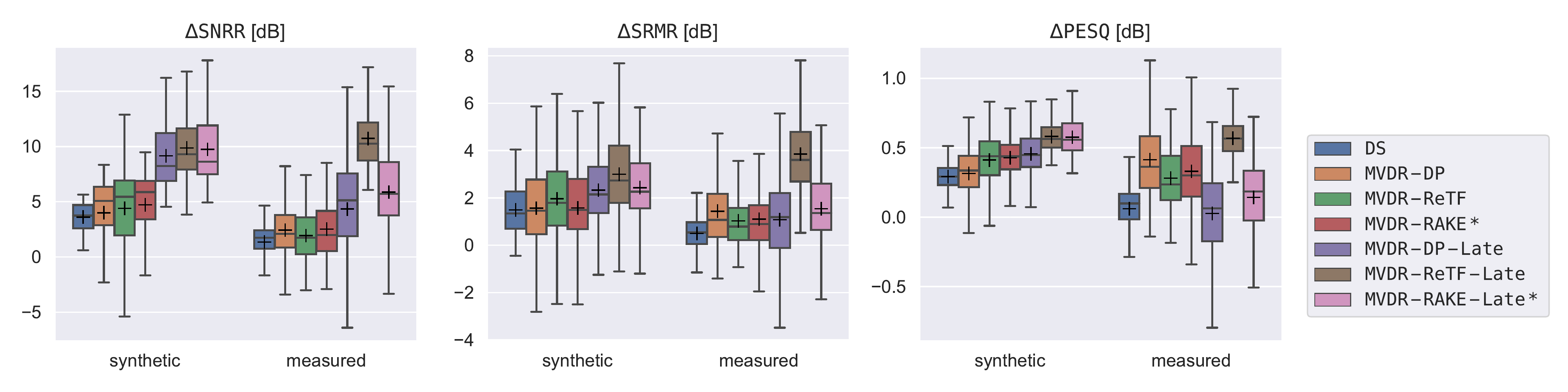}
    \caption{
    Boxplot showing the comparison of different echo-agnostic and echo-aware (*) beamformers for the room configuration $\mathtt{011111}$ ($\RTsixty \approx 600 $ ms) on measured and synthetic data  for all combinations of source-array positions in the \dEchorate~dataset.
    \\\protect Mean values is indicated as $+$, while whiskers indicates extreme values.}
    \label{fig:bf_boxplot}
\end{figure*}

Numerical results are reported in \Cref{fig:bf_boxplot}.
On synthetic data, as expected, one can see that the more information is used, the better performances are. Including late reverberation statistics considerably boosts performance in all cases.
Both the \acp{ReTF}-based and the echo-aware beamformers significantly outperform the simple designs based on direct path only. While the two designs perform comparably in terms of \ac{iSNRR} and \ac{iPESQ}, the former has a slight edge over the latter in terms of median \ac{iSRMR}.
A possible explanation is that \ac{GEVD} methods tend to consider the stronger and more stable components of the \acp{ReTF}, which in the considered scenarios may identify with the earlier portion of the \acp{RIR}.
Moreover, since it is not constrained by a fixed echo model, the \acp{ReTF} can capture more information, \eg, frequency-dependent attenuation coefficients. 
Finally, one should consider the compacity of the model~(\ref{sec:appl:echomodel}) with respect to the \ac{ReTF} model in terms of the number of parameters to be estimated. In fact, when considering 4 echoes, only 8 parameters per channel are needed, as opposed to several hundreds for the \ac{ReTF} (ideally, as many as the number of frequency bins per channel).

When it comes to measured \acp{RIR}, however, the trends are different.
Here, the errors in echo timings due to calibration mismatch and the richness of real acoustic propagation lead to a drop in performance for echo-aware methods, both in terms of means and variances.
This is clearest when considering the \ac{iPESQ} metric, which also accounts for artifacts.
the echo-agnostic beamformer considering late reverberation $\MVDRretfLate$ outperforms the other methods, maintaining the trend exhibited on simulated data.
Finally, conversely to the $\MVDRretfLate$, the $\MVDRrakeLate$ yields a significant portion of negative performances.
As already observed in~\cite{kowalczyk2019raking}, this is probably due to the tiny annotation mismatches in echo timings as well as the fact that their frequency-dependent strengths, induced by reflective surfaces, are not modeled in rake beamformers.
This suggests that in order to be applicable to real conditions, future work in echo-aware beamforming should include finer blind estimates of early echo properties from signals, as investigated in, \eg, \cite{tukuljac2018mulan,dicarlo2020blaster}.

\subsection{Application: Room Geometry Estimation}
\begin{figure}[h]
    \includegraphics[width=0.95\linewidth]{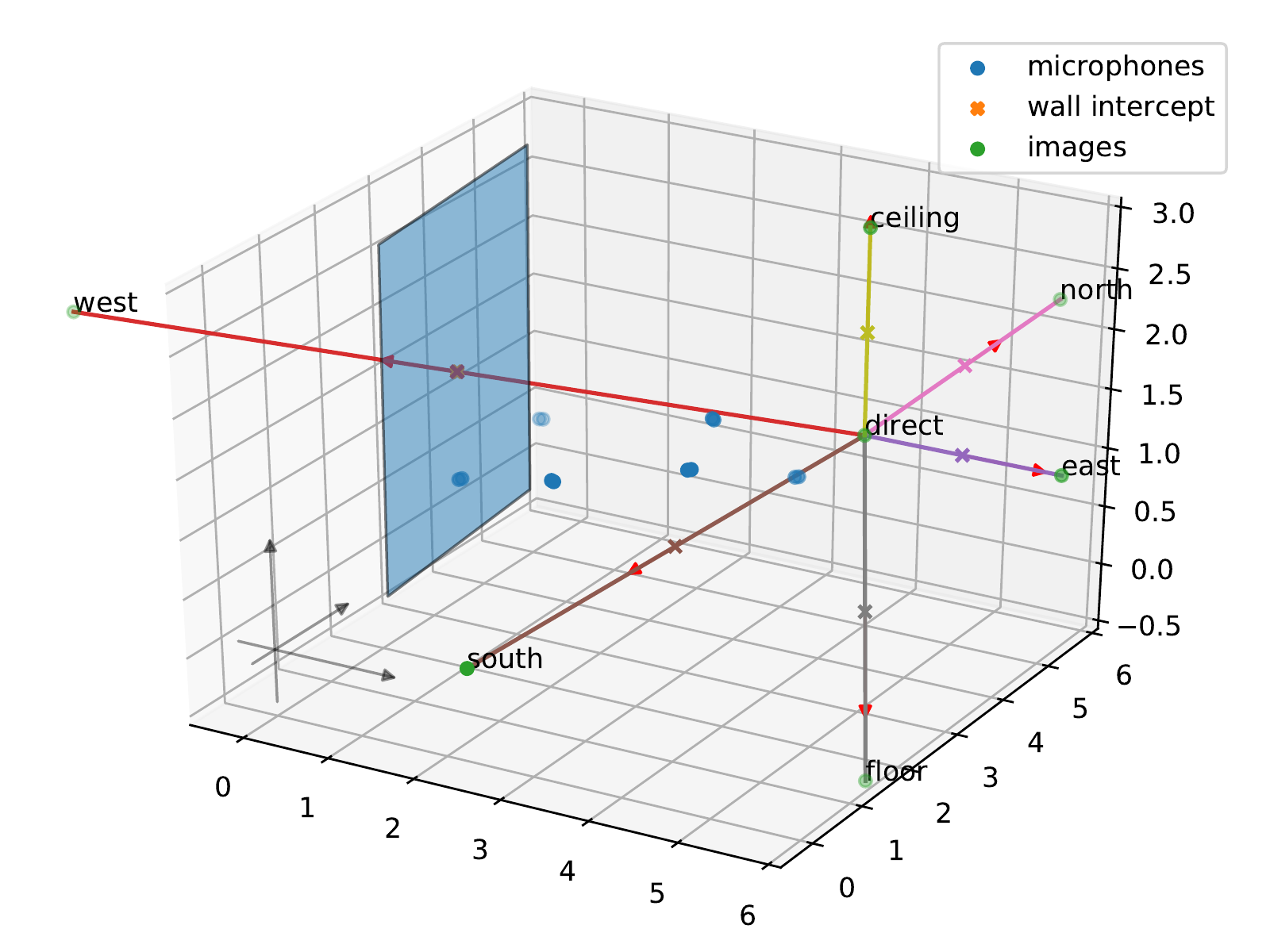}
    \caption{Images source estimation and reflector estimation for one of the sound sources in the dataset.}
    \label{fig:wall_rec}
\end{figure}

The shape of a convex room can be estimated knowing the positions of first-order image sources.
Several methods have been proposed which take into account different levels of prior information and noise (see~\cite{remaggi2017acoustic,crocco2018room} for a review).
When the echoes' \ac{TOA} and their labeling are known for 4 non-coplanar microphones, one can perform this task using simple geometrical reasoning as in \cite{dokmanic2013acoustic}.
In details, the 3D coordinates of each image source can be retrieved solving a multilateration problem ~\cite{beck2008ExactProblems}, namely the extension of the trilateration problem to 3D space, where the goal is to estimate the relative position of an object based on the measurement of it distance with respect to anchor points.
Finally, the position and orientation of each room facet can be easily derived from the \ac{ISM} equations as the plane bisecting the line joining the real source position and the position of its corresponding image (see \Cref{fig:wall_rec})

In \dEchorate{}, the annotation of all the first order echo timings are available, as well as correspondences between echoes and room facets. This information can be used directly as input for the above-mentioned multilateration algorithm.
We illustrate the validity of these annotations by employing the \ac{RooGE} technique in \cite{dokmanic2013acoustic} (with know labels) based on them.

\begin{table*}[h!]
\centering
\caption{\label{tab:res_rooge} \acf{DE} in centimeters and \acf{AE} in degrees between ground truth and estimated room facets using each of the sound sources ($\#1$ to $\#4$) as a probe. For each wall, bold font is used for the source yielding the best \ac{DE} and \ac{AE}, while italic highlights outliers when present.}

\begin{tabular}{c|cc|cc|cc|cc}
\toprule
source id &	1	& &	2	& &	3	& &	4 &	\\
wall &	DE&	AE&	DE&	AE&	DE&	AE&	DE&	AE\\
\hline
west &	0.74	& 8.99$\ang{}$      & 4.59	& 8.32$\ang{}$  & 5.89	& 5.75$\ang{}$	& $\mathbf{0.05}$    & $\mathbf{2.40\ang{}}$\\
east &	$\mathbf{0.81}$	& $\mathbf{0.08\ang{}}$      & 0.9	& 0.50$\ang{}$	&$\mathit{69.51}$	& $\mathit{55.7\ang{0}}$	& 0.31    & 0.21$\ang{}$\\
south&	3.94	&16.0$\ang{8}$      & $\mathbf{0.18}$	& 1.77$\ang{}$	&$\mathit{14.37}$ & $\mathit{18.5\ang{5}}$	& 0.82    & $\mathbf{1.65\ang{}}$\\
north&	1.34	& 0.76$\ang{}$	    & 1.40	& 8.94$\ang{}$	& $\mathbf{0.63}$	& $\mathbf{0.17\ang{}}$	& 2.08    & 1.38$\ang{}$\\
floor&	 $\mathbf{5.19}$	& $\mathbf1.76{\ang{}}$	    & 7.27	& 2.66$\ang{}$	& 7.11	& 2.02$\ang{}$	& 5.22    & 1.90$\ang{}$\\
ceiling&1.16	& 0.28$\ang{}$	    & 0.67	& 0.76$\ang{}$	& $\mathbf{0.24}$	& 1.16$\ang{}$	& 0.48    & $\mathbf{0.26\ang{}}$\\

\bottomrule
\end{tabular}
\end{table*}

\Cref{tab:res_rooge} shows the results of the estimation of the room facets position in terms of \acf{DE} (in centimeters) and surface orientation error, (dubbed here \acf{AE}, in degrees) using a single source and all 30 microphones, namely the 6 arrays.
Room facets are estimated using each of the sources $\#1$ to $\#4$ as a probe.
Despite a few outliers, the majority of facets are estimated correctly in terms of their placement and orientation with respect to the coordinate system computed in \Cref{subsec:annotation}. For instance, using source $\#4$, all 6 surfaces were localized with 1.49 cm \ac{DE} on average and their inclinations with $1.3\ang{}$ \ac{AE} on average.
Apart from the outliers, these results are in line with the ones reported by Dokmani\'c \etal in the work~\cite{dokmanic2013acoustic} using a setup of 5 microphones listening to 1 sound source.
\\Furthermore, one can use all the 4 sources to estimate the room geometry as suggested in~\cite{crocco2017uncalibrated}. By doing so, the entire room geometry estimation results in 1.15 cm \ac{DE} and $2.6\ang{}$ \ac{AE} on average.

The small errors are due to a concurrency of multiple factors, such as tiny offsets in the annotations and the ideal shoebox approximation.
In the real recording room, some gaps were present between revolving panels in the room facet.
In addition, it is possible that for some (image source, receiver) pairs the far-field assumption is not verified, causing inaccuracies when  inverting the \ac{ISM}.
The 2 outliers for source $\#3$ are due to a wrong annotation caused by the source directivity which induced an echo mislabeling.
When a wall is right behind a source, the energy of the related $1^\text{st}$ reflection is very small and might not appear in the \acp{RIR}.
This happened for the eastern wall and a second order image was taken instead.
Finally, the contribution of multiple reflections arriving at the same time can result in large late spikes in estimated \acp{RIR}.
This effect is particularly amplified when the microphone and loudspeakers exhibit long impulse responses.
As a consequence, some spikes can be miss-classified.
This happened for the southern-wall where again a second-order image was taken instead.
Note that such echo mislabelings can either be corrected manually or using Euclidean distance matrix criteria as proposed in \cite{dokmanic2013acoustic}. Overall, this experiment illustrates well the interesting challenge of estimating and exploiting acoustic echoes in \acp{RIR} when typical sources and receivers with imperfect characteristics are used.

% \begin{figure*}[h]
%     \subfigure{
%         \includegraphics[width=0.49\textwidth]{figures/estimated_image}}
%     \hfill
%     \subfigure{
%         \includegraphics[width=0.49\textwidth]{figures/estimated_reflector}}

%     \caption{Images source estimation (right) and corresponding reflector estimation (left) for one of the sound sources in the dataset.}
%     \label{fig:wall_rec}
% \end{figure*}
ec
\section{Conclusions and Perspectives}\label{sec:conclusion}
This paper introduced a new database of room impulse responses featuring accurate annotation of early echo timings that are consistent with source, microphone and room facet positions.
These data can be used to test methods in the room geometry estimation pipeline and in echo-aware audio signal processing.
In particular, robustness of these methods can be validated against different levels of \RTsixty, \SNR{}, surface reflectivity, proximity, or early echo density.

This dataset paves the way to a number of interesting future research directions.
By making this dataset freely available to the audio signal processing community, we hope to foster research in \ac{AER} and echo-aware signal processing in order to improve the performance of existing methods on real data.
Moreover, the dataset could be updated by including more robust annotations derived from more advanced algorithms for calibration and \ac{AER}. 

In addition, the data analysis conducted in this work brings the attention to exploring the impact of mismatch between simulated and real \acp{RIR} on audio signal processing methods. Finally, by using the pairs of simulated vs. real \acp{RIR} available in the dataset, it should be possible to develop techniques to convert one to the other, using style transfer or domain adaptation techniques, thus opening the way to new types of learning-based acoustic simulators.

\section*{Appendix}
\subsection*{Room materials}\label{app:materials}

\begin{table}[h]
    \centering
    \caption{\label{tab:materials} Materials covering the acoustic laboratory in Bar-Ilan University.}
    \begin{tabular}{c|cl}
        \toprule
         Surface &  Mode &  Material  \\
         \midrule
         Floor &  absorbent & Hairy carpet   \\
         Ceiling & absorbent &  Glass wool mats covered with porous tin \\
         Ceiling & reflective & Formica (20 mm thick)\\
         Walls & absorbent & Glass wool mats covered with porous tin \\
         Walls & reflective & Panels: Formica (20 mm thick) \\
         {} & {} & Wall: Plaster \\
         \bottomrule
    \end{tabular}
\end{table}

%%%%%%%%%%%%%%%%%%%%%%%%%%%%%%%%%%%%%%%%%%%%%%
%%                                          %%
%% Backmatter begins here                   %%
%%                                          %%
%%%%%%%%%%%%%%%%%%%%%%%%%%%%%%%%%%%%%%%%%%%%%%

\begin{backmatter}

\section*{Acknowledgements}%% if any
Luca Remaggi, Marco Crocco, Alessio Del Bue, and Robin Scheibler are thanked for help during experimental design.

\section*{Availability of data and materials}%% if any
The database is publicly available at \linkData{} and \linkCode{}.

\section*{Abbreviations}%% Mandatory
\begin{acronym}[UMLX]
  \acro{AE}{Angular Error}
  \acro{AER}{Acoustic Echo Retrieval}
  \acro{ASR}{Automatic Speech Recognition}
  \acro{DE}{Distance Error}
  \acro{DER}{Direct-to-Early Ratio}
  \acro{DRR}{Direct-to-Reverberant ratio}
  \acro{DOA}{Direction of Arrival}
  \acro{ESS}{Exponentially Swept-frequency Sine}
  \acro{GEVD}{Generalized Eigenvector Decomposition}
  \acro{GoM}{Goodness of Match}
  \acro{MDS}{Multi-Dimensional Scaling}
  \acro{MVDR}{Minimum Variance Distortionless Response}
  \acro{nULA}{non-Uniform Linear Array}
  \acro{PESQ}{Perceptual Evalution of Speech Quality}
  \acro{RIR}{Room Impulse Response}
  \acro{ReTF}{Relative Transfer Function}
%   \acro{RT}{Reverberation Time}
  \acro{TOA}{Time of Arrival}
  \acro{TDOA}{Time Difference of Arrival}
  \acro{ISM}{Image Source Method}
  \acro{SE}{Speech Enhancement}
%   \acro{SNR}{Signal-to-Noise Ratio}
  \acro{SNRR}{Signal-to-Noise plus Reverberation Ratio}
  \acro{iPESQ}{Perceptual Evaluation of Speech Quality improvement}
  \acro{iSNRR}{Signal-to-Noise plus Reverberation Ratio improvement}
  \acro{iSRMR}{Speech-to-Reverberation energy Modulation Ratio improvement}
  \acro{RooGE}{Room Geometry Estimation}
  \acro{WSJ}{Wall Street Journal}
\end{acronym}

% \section*{Ethics approval and consent to participate}%% if any
% Text for this section\ldots

% \section*{Competing interests}
% The authors declare that they have no competing interests.

% \section*{Consent for publication}%% if any
% Text for this section\ldots

% \section*{Authors' contributions}
% Text for this section \ldots

% \section*{Authors' information}%% if any
% Text for this section\ldots

%%  After submission of the .TEX file,                     %%
%%  you will be prompted to submit your .BBL file.         %%

% if your bibliography is in bibtex format, use those commands:
\bibliographystyle{bmc-mathphys} % Style BST file (bmc-mathphys, vancouver, spbasic).
\bibliography{ref_dechorate}

\end{backmatter}
\end{document}